%% file: main.tex
\documentclass[letterpaper, 10 pt, conference]{IEEEtran} 

\usepackage{graphicx}
\usepackage{times}
\usepackage{amsmath,amsbsy,amssymb,mathtools}
\usepackage{color}
\usepackage[us]{datetime}
\usepackage{url}
\usepackage{theorem}
\usepackage{multirow}
\usepackage{footnote}
\usepackage{nomencl}
\usepackage{filecontents}
\usepackage{cite}
\usepackage{algorithm}
\usepackage{algpseudocode}
\usepackage{stackengine}

\usepackage{caption}
\usepackage{subcaption}
\usepackage{commath}

\newcommand{\R}{\mathbb{R}} 

\theoremstyle{plain}
\newtheorem{theorem}{Theorem}

\newtheorem{lemma}{Lemma}
\newtheorem{corollary}{Corollary}
\newtheorem{remark}{Remark}

\newtheorem{assumption}{Assumption}
\newtheorem{cvxCondition}{Condition}
\usepackage{lipsum}
\newcommand{\qed}{\hfill\rule{1.5ex}{1.5ex}}
\newcommand{\qedd}{\tag*{$\blacksquare$}}

\algnewcommand\algorithmicswitch{\textbf{switch}}
\algnewcommand\algorithmiccase{\textbf{case}}
\algdef{SE}[SWITCH]{Switch}{EndSwitch}[1]{\algorithmicswitch\ #1\ \algorithmicdo}{\algorithmicend\ \algorithmicswitch}%
\algdef{SE}[CASE]{Case}{EndCase}[1]{\algorithmiccase\ #1}{\algorithmicend\ \algorithmiccase}%
\algtext*{EndSwitch}%
\algtext*{EndCase}%

\IEEEoverridecommandlockouts                              

\title{\large \bf
A Computationally Efficient Hamilton-Jacobi-based Formula for State-Constrained Optimal Control Problems
}

\author{Donggun Lee and Claire J. Tomlin
\thanks{This research is supported by ONR under the BRC program in multibody control systems, by DARPA under the Assured Autonomy program, and by
NSF grant \#1837244.}
\thanks{Donggun Lee is with the Department of Mechanical Engineering, University of California, Berkeley, USA.
        {\tt\small donggun\_lee@berkeley.edu}}%
\thanks{Claire J. Tomlin is with the Department of Electrical Engineering and Computer Sciences, University of California, Berkeley, USA.
        {\tt\small tomlin@eecs.berkeley.edu}}%
}

\renewcommand{\baselinestretch}{.93}
\begin{document}


\onecolumn
\copyright 2021 IEEE. Personal use of this material is permitted. Permission from IEEE must be obtained for all other uses, in any current or future media, including reprinting/republishing this material for advertising or promotional purposes, creating new collective works, for resale or redistribution to servers or lists, or reuse of any copyrighted component of this work in other works

This work has been submitted to the IEEE for possible publication. Copyright may be transferred without notice, after which this version may no longer be accessible.
\twocolumn

\newpage

\maketitle
\thispagestyle{plain}
\pagestyle{plain}

\begin{abstract}
This paper investigates a Hamilton-Jacobi (HJ) analysis to solve finite-horizon optimal control problems for high-dimensional systems.
Although grid-based methods, such as the level-set method \cite{Osher02}, numerically solve a general class of HJ partial differential equations, the computational complexity is exponential in the dimension of the continuous state.
To manage this computational complexity, methods based on Lax-Hopf theory have been developed for the state-unconstrained optimal control problem under certain assumptions, such as affine dynamics and state-independent stage cost.
Based on the Lax formula \cite{evans10}, this paper proposes an HJ formula for the state-constrained optimal control problem for nonlinear systems.
We call this formula \textit{the generalized Lax formula} for the optimal control problem. 
The HJ formula provides both the optimal cost and an optimal control signal.
We also provide an efficient computational method for a class of problems for which the dynamics is affine in the state, and for which the stage and terminal cost, as well as the state constraints, are convex in the state.
This class of problems does not require affine dynamics and convex stage cost in the control.
This paper also provides three practical examples.
\end{abstract}

\section{Introduction}

Hamilton-Jacobi (HJ) analysis is a method for solving optimal control and differential game problems, by formulating an HJ partial differential equation (PDE) which encodes the dynamics and the cost \cite{evans10,evans1984differential,mitchell2005time,margellos2011hamilton,fisac2015reach,altarovici2013general}.
It has been widely utilized in a variety of fields, including 
autonomous driving \cite{chen2015safe,chen2017reachability}, air traffic \cite{tomlin1998conflict,parzani2017hamilton}, robotics \cite{fridovich2018}, economics \cite{achdou2017income}, and finance \cite{rochet1985taxation,forsyth2007numerical}.

In this paper, we are concerned with solving a finite-horizon optimal control problem; specifically, we want to find an optimal state trajectory and a control signal which minimize a cost while satisfying a state constraint.
The cost consists of the integration of the stage cost and the terminal cost.
For the state-unconstrained and state-constrained optimal control problems, the corresponding HJ equations are presented in \cite{evans10} and \cite{altarovici2013general}, respectively.
These HJ equations can be solved by grid-based methods, such as the level-set method \cite{Osher02} and fast marching method \cite{sethian1996fast}.
Unfortunately, the grid-based methods lead to computational complexity exponential in the dimension of the continuous state, which makes it unrealistic to solve HJ equations of state dimension more than five or six.


To alleviate computational complexity, various methods have been developed.
For instance, system decomposition methods \cite{chen2018decomposition,chen2017exact}, projection methods \cite{Mitchell03}, HJ PDE decomposition methods \cite{mitchell2011scalable,donggun2019}, set-based approximations \cite{Stipanovic2003,Hwang2005}, and learning-based methods \cite{rubies2019classification} have been proposed.
The system decomposition \cite{chen2018decomposition,chen2017exact} and projection methods \cite{Mitchell03} decompose the problem into multiple lower-dimensional problems and provide an approximation (and sometimes exact) solution by combining the solutions of the lower-dimensional problems.
On the other hand, the HJ PDE decomposition methods \cite{mitchell2011scalable,donggun2019} formulate an HJ PDE in a lower-dimensional space.
For reachability problems, set-based approximation methods \cite{Stipanovic2003,Hwang2005} utilize ellipsoidal or polytopic representation to approximate the subzero-level sets of the solution to the HJ PDE.
In the field of learning, \cite{rubies2019classification} utilizes neural networks to approximate the optimal control as a binary classifier, and \cite{djeridane2006neural} utilizes a feed-forward neural network to approximate the HJ PDE's solution to a parameterized function.
Recently, the connections between neural networks and viscosity solutions to HJ PDEs have been mathematically investigated \cite{darbon2020some}.

Lax-Hopf theory is an alternative method for solving high-dimensional HJ PDEs, by formulating an optimization problem, which can be solved numerically using temporal discretization and gradient-based optimization techniques \cite{evans10,hopf1965generalized}.
If the temporally discretized problem is convex, the gradient-based optimization techniques provide an optimal solution with substantial computational savings. 

In Lax-Hopf theory, two types of formulae have been developed: Lax-based and Hopf-based formulae.
The Lax-based formulae provide an optimal control problem in which the control is specified when the Hamiltonian is convex in the costate, and the Hopf-based formulae provide an optimal control problem in which the costate is specified when the given terminal value function is convex in the state.

As the first Lax-based and Hopf-based formulae, Lax and Oleinik have contributed to the Hopf-Lax formula \cite{evans10} and Hopf proposed the Hopf formula \cite{hopf1965generalized}, respectively.
These two formulae assume that the Hamiltonian depends on the costate but not on the time and the state.
This implies that the dynamics and the stage cost depend on the control but not on the time and the state.
Later, by relaxing the assumption of the Hopf formula \cite{hopf1965generalized}, the generalized Hopf formula \cite{lions1986hopf} was proposed, which allows time dependency in the Hamiltonian, the dynamics, and the stage cost.
Based on this, Darbon et al. \cite{darbon2016algorithms} have proposed a Hopf-based method for linear dynamics and state-independent stage cost.
Thus, the state-of-the-art Lax-Hopf theory assumes 1) linear dynamics, 2) state-independent stage cost, and 3) no state constraint for solving optimal control or dynamic game problems.
In recent development, for nonlinear dynamics and general stage cost that depends on the time, state, and control, conjectures for both Lax-based and Hopf-based formulae \cite{chow2019algorithm} have been presented.



\subsection{Contribution}

This paper proposes a Lax-based formula for the finite-horizon state-constrained optimal control problem where the assumption for the Lax-based formulae holds: the corresponding Hamiltonian is convex in the costate \cite{altarovici2013general}.
Our method aims to deal with nonlinear dynamics and general stage cost which depends on the time, state, and control.
In this work, we build on the Hopf-Lax formula \cite{evans10,bardi1984hopf}.

Our earlier work \cite{lee2020hopf} presented  a Lax formula for reach-avoid problems in which an optimal control signal and an optimal time-to-reach are determined, to drive the system from an initial state to a goal, while avoiding unsafe sets.
In \cite{lee2020hopf}, the reach-avoid problem is converted to a state-constrained optimal control problem. 
Theorem \ref{thm:OptCtrl_PostProcess} in this current paper is utilized to derive an optimal control signal for the reach-avoid problem in \cite{lee2020hopf}; the proof details of Theorem \ref{thm:OptCtrl_PostProcess} are presented in the current paper.

The contributions of this paper are 1) the proposal of a Lax formula for the finite-horizon state-constrained optimal control problem, 2) a numerical algorithm to compute an optimal state trajectory and control signal, 3) proof for significant computational savings under following conditions: (a) the dynamics is affine in the state; (b) the stage and terminal cost, as well as the state constraints, are convex in the state, and 4) 
the provision of three numerical examples.

\subsection{Organization}
The organization of this paper is as follows. 
In Section \ref{sec:background}, we present the problem description. 
In Section \ref{sec:NewFormulation}, we propose a Lax-based formula for the finite-horizon state-constrained optimal control problem and present a numerical algorithm to compute an optimal state trajectory and control signal.
In Section \ref{sec:CvxAnalysis_GeneralizedHopfLax}, using convexity analysis, we present the conditions under which our method (the generalized Lax formula) provides efficient computation.
Section \ref{sec:Examples} provides three numerical examples.
In Section \ref{sec:conclusion}, we present conclusions and some potential applications for future work.

In terms of notation in this paper, we use $a^*$ for the Legendre-Fenchel transformation (convex conjugate) of $a$, and $a_*$ to represent the optimal value of $a$.

\section{State-Constrained Optimal Control Problem}
\label{sec:background}

Given initial time and state $(t,x)$, we would like to solve the finite-horizon state-constrained optimal control problem for the dynamic system:
\begin{align}
    \vartheta (t,x) \coloneqq \lim_{\epsilon\rightarrow 0} \vartheta^\epsilon (t,x),
    \label{eq:def_vartheta}
\end{align} 
where
\begin{align}
    & \vartheta^\epsilon (t,x) \coloneqq \inf_{\alpha} \int_t^T L(s,\mathrm{x}(s),\alpha(s))ds + g(\mathrm{x}(T)) 
    \label{eq:def_vartheta_ep}\\
    & \text{subject to }
    \begin{cases}
        \dot{\mathrm{x}}(s) = f(s,\mathrm{x}(s),\alpha(s)), & s\in [t,T], \\
        \mathrm{x}(t) = x,\\
        \alpha(s)\in A, & s\in [t,T],\\
        c(s,\mathrm{x}(s)) \leq \epsilon, & s\in [t,T].
    \end{cases}
    \label{eq:def_vartheta_ep_const}
\end{align}
Here, $L:[t,T]\times\R^n\times A\rightarrow \R$ is the stage cost, $g:\R^n\rightarrow\R$ is the terminal cost, $f:[t,T]\times\R^n\times A \rightarrow\R^n$ is the system dynamics, $c:[t,T]\times\R^n\rightarrow\R$ is the state constraint, and $\alpha\in \mathcal{A}(t)$ is the control signal where $\mathcal{A}(t)$ is the set of admissible control signals:
\begin{align}
    \mathcal{A}(t) \coloneqq \{\alpha:[t,T]\rightarrow A ~|~ \|\alpha\|_{L^\infty(t,T)} <\infty\},
    \label{eq:def_CtrlTraj}
\end{align}
and $A$ is a compact subset in $\R^m$. 
In practice, $c(s,\cdot)$ represents unsafe regions at $s\in[t,T]$ so that $c(s,x)<0$ for $x\in\R^n$ away from the unsafe regions, $c(s,x)=0$ for $x$ on the boundary of the unsafe regions, and $c(s,x)>0$ for $x$ in the unsafe regions.

This paper solves problem \eqref{eq:def_vartheta}. Note that, in general, 
\begin{align}
    \vartheta(t,x) \leq  \vartheta^0 (t,x),
    \label{eq:property_vartheta}
\end{align}
For example, consider the state-constrained optimal control problem with the initial time and state, $(t,x)=(0,0)\in\R\times\R$,
\begin{align}
    & \vartheta^\epsilon (0,0) = \inf_{\alpha} \int_0^1 | \mathrm{x}(s) | ds  
    \notag\\
    & \text{subject to }
    \begin{cases}
        \dot{\mathrm{x}}(s) = \alpha(s), & s\in [0,1], \\
        \mathrm{x}(0) = 0,\\
        \alpha(s)\in \{-1,1\}, & s\in [0,1],\\
        | \mathrm{x}(s)| \leq \epsilon, & s\in [0,1].
    \end{cases}
    \notag
\end{align}
For any $\epsilon>0$, $0< \vartheta^\epsilon(0,0)\leq \epsilon$ since there exists $\alpha$ such that $|\mathrm{x}(s)|\leq \epsilon$ for all $s\in[0,1]$.
However, $\vartheta^0 (0,0)=\infty$ since there is no admissible control signal that satisfies the state constraint, $|\mathrm{x}(s)|\leq 0$ for all $s\in[0,1]$. 
Thus,
\begin{align*}
    \vartheta(0,0)=\lim_{\epsilon\rightarrow0}\vartheta^\epsilon(0,0)=0 < \vartheta^0 (t,x)= \infty.
\end{align*}
The equality in \eqref{eq:property_vartheta} holds when $A$ is convex  \cite{altarovici2013general}. 
Although $\vartheta^\epsilon(t,x)$ allows $\epsilon$-violation to the state constraint ($c(s,\mathrm{x}(s))\leq\epsilon$), $\epsilon$ converges to 0.
\cite{altarovici2013general} solves $\vartheta^0(t,x)$ under the assumption that the control constraint ($A$) is convex; in the current paper we relax this assumption.

In this paper, we assume the following.
\begin{assumption}[Lipschitz continuity and compactness]
~
\begin{enumerate}
    \item the control set $A$ is compact;
    
    \item $f:[0,T]\times\R^n\times A\rightarrow\R^n$ is Lipschitz continuous in $(s,x)$ for each $a\in A$:
    \begin{align}
    \begin{split}
        \|f(s_1,x_1,a)  - & f(s_2,x_2,a)\|\leq \\
        &L_f (|s_1-s_2|+\|x_1-x_2\|);
    \end{split}
    \end{align}
    
    \item the stage cost $L:[0,T]\times\R^n\times A\rightarrow\R$ is Lipschitz continuous in $(s,x)$ for each $a\in A$:
    \begin{align}
    \begin{split}
        \|L(s_1,x_1,a)  - & L(s_2,x_2,a)\|\leq \\
        &L_L (|s_1-s_2|+\|x_1-x_2\|);
    \end{split}
    \end{align}

    \item the terminal cost $g:\R^n\rightarrow\R$ is Lipschitz continuous in $x$:
    \begin{align}
        \|g(x_1)-g(x_2)\| \leq L_g \|x_1-x_2\|;
    \end{align}
    
    \item the state constraint $c:[0,T]\times\R^n \rightarrow\R$ is Lipschitz continuous in $(s,x)$:
    \begin{align}
    \begin{split}
        \|c(s_1,x_1)-&c(s_2,x_2)\| \leq \\
         &L_c (|s_1-s_2|+\|x_1-x_2\|).
     \end{split}
         \label{eq:Lipschitz_stateConst}
    \end{align}
    
    \item the stage cost ($L$) and the terminal cost ($g$) are bounded below: there exists $C\in\R$ such that
    \begin{align}
        L(s,x,a) \geq C, \quad  g(x) \geq C
        \label{eq:assump_boundness}
    \end{align}
    for all $(s,x,a)\in[0,T]\times\R^n\times A$.
\end{enumerate}
\label{assum:BigAssum}
\end{assumption}
Under Assumption \ref{assum:BigAssum}, we first show the existence of $\vartheta(t,x)$ in \eqref{eq:def_vartheta}.
\begin{lemma}
    Suppose Assumption \ref{assum:BigAssum} holds,
    $\vartheta(t,x):[0,T]\times\R^n\rightarrow\R\cup\{\infty\}$ exists. 
\end{lemma}
\textbf{Proof.} If $\epsilon_1 > \epsilon_2 >0$, $\vartheta^{\epsilon_1}(t,x)\leq \vartheta^{\epsilon_2}(t,x)$. Thus, the limit of $\vartheta^\epsilon$ always exists in $\R\cup\{\infty\}$. 
$\qed$

\section{The Generalized Lax Formula and Optimal Control}
\label{sec:NewFormulation}

In this section, we propose a Lax-based formula for the state-constrained optimal control problem in \eqref{eq:def_vartheta}. 
This is derived in three steps: 1) in Section \ref{sec:NewFormulation1}, we first derive a generalized Lax formula for the state-unconstrained problem that does not contain the state constraint $c(s,\mathrm{x}(s))$ in \eqref{eq:def_vartheta_ep_const}; 2) in Section \ref{sec:OptCtrl_NoStateConst}, we investigate the relationship between control signals for the given problem \eqref{eq:def_vartheta} and the generalized Lax formula proposed in Section \ref{sec:NewFormulation1}; and 3) using Section \ref{sec:NewFormulation1} and \ref{sec:OptCtrl_NoStateConst}, the generalized Lax formula for the state-constrained optimal control problem \eqref{eq:def_vartheta} is derived in Section \ref{sec:NewFormulation_StateConst}.

\subsection{The generalized Lax formula for the state-unconstrained optimal control problem}
\label{sec:NewFormulation1}

In this subsection, we propose the generalized Lax formula in Theorem \ref{thm:GHopfLax} for the state-unconstrained problem: solving \eqref{eq:def_vartheta} without the state constraint $c(s,\mathrm{x}(s))$ in \eqref{eq:def_vartheta_ep_const}:
\begin{align}
    \vartheta(t,x) = \inf_{\alpha}\int_t^T L(s,\mathrm{x}(s),\alpha(s))dt+g(\mathrm{x}(T)),\label{eq:NoConst_vartheta_cost}\\
    \text{subject to }\begin{cases}\dot{\mathrm{x}}(s)=f(s,\mathrm{x}(s),\alpha(s)),&s\in[t,T],\\ \mathrm{x}(t)=x,&\\\alpha(s)\in A,&s\in[t,T].\end{cases}
    \label{eq:NoConst_vartheta_const}
\end{align}
Note that here we simply use $\vartheta(t,x)$, as $\epsilon$ is only relevant for the state-constrained problem.

We first show, in Theorem \ref{thm:HJB_PDE} from \cite{evans10}, that $\vartheta$ solves the Hamilton-Jacobi-Bellman (HJB) PDE.
\begin{theorem}[HJB PDE \cite{evans10}]
    Suppose Assumption \ref{assum:BigAssum} holds.
    For $(t,x)\in[0,T]\times\R^n$, $\vartheta$ in \eqref{eq:NoConst_vartheta_cost} subject to \eqref{eq:NoConst_vartheta_const} is the unique viscosity solution to
    \begin{align}
        \vartheta_t (t,x) - H(t,x,D\vartheta(t,x))=0 \quad &\text{in}\quad (0,T)\times\R^n,\\
        \vartheta (T,x) = g(x)  \quad &\text{on}\quad  \{t=T\}\times\R^n,
    \end{align}
    where $H:[0,T]\times \R^n\times \R^n \rightarrow \R$
    \begin{align}
        H(t,x,p) \coloneqq \max_{a\in A}-p\cdot f(t,x,a) - L(t,x,a).
        \label{eq:Hamiltonian}
    \end{align}
    Denote that $\vartheta_t = \frac{\partial \vartheta}{\partial t}$ and $D\vartheta = \frac{\partial \vartheta}{\partial x}$.
    \label{thm:HJB_PDE}
\end{theorem}

We consider a control space transformation: for $(s,x,a)\in[t,T]\times\R^n\times A$,
\begin{equation}
    b = -f(s,x,a) \in \R^n ,
    \label{eq:ctrlSpace_Transformation}
\end{equation}
where $f$ and $A$ are the dynamics and the control constraint of the system, respectively.
For $(s,x)$, the control constraint for $b$ is denoted by
\begin{align}
    B(s,x) := \{ -f(s,x,a) ~|~ a\in A  \}.
    \label{eq:Trans_CtrlSet}
\end{align}
$B(s,x)$ is compact for all $(s,x)$ since $A$ is compact and $f$ is Lipschitz as in Assumption \ref{assum:BigAssum}.
By the control space transformation in \eqref{eq:ctrlSpace_Transformation}, for any state trajectory $\mathrm{x}$ solving \eqref{eq:NoConst_vartheta_const}, there exists a control signal $\beta:[t,T]\rightarrow\R^n$ such that $\mathrm{x}$ solves
\begin{align}
\begin{array}{l}  
\dot{\mathrm{x}}(s)=-\beta(s),\quad s\in[t,T], \\ [1ex] 
\mathrm{x}(t)=x,\\[1ex] 
\beta(s)\in B(s,\mathrm{x}(s)), \quad s\in[t,T].
\end{array}  
    \label{eq:newDyn}
\end{align}
With respect to $\mathrm{x}$ solving \eqref{eq:newDyn}, $\vartheta$ in \eqref{eq:NoConst_vartheta_cost} subject to \eqref{eq:NoConst_vartheta_const} is converted to
\begin{align}
    \vartheta(t,x) = \inf_{\beta}\int_t^T L^b (s,\mathrm{x}(s),\beta(s)) ds + g(\mathrm{x}(T)),
    \label{eq:ValueFunction_Def_CtrlTrans}\\
    \text{subject to} 
    \begin{cases}
        \dot{\mathrm{x}}(s) = -\beta(s), & s\in [t,T], \\
        \mathrm{x}(t) = x,\\
        \beta(s) \in B(s,\mathrm{x}(s)), & s\in [t,T],\\
        \end{cases} \quad
    \label{eq:ValueFunction_Constraint_CtrlTrans}
\end{align}
where 
\begin{align}
    L^b (s,x,b) := \min_{a\in A} L(s,x,a) \quad \text{s.t.}\quad f(s,x,a)=-b.
    \label{eq:Def_Lv}
\end{align}
$L^b$ in \eqref{eq:Def_Lv} and $H$ in \eqref{eq:Hamiltonian} satisfy the following properties.

\begin{lemma} Suppose Assumption \ref{assum:BigAssum} holds.
$L^b$ in \eqref{eq:Def_Lv} and $H$ in \eqref{eq:Hamiltonian} have the following properties.
    \begin{align}
        &(L^b)^* (s,x,p)= H (s,x,p) ~~~~~ \text{in }[0,T]\times\R^n\times\R^n, \label{eq:property_ConvConj0}\\
        &(L^b)^{**}(s,x,b) = H^*(s,x,b) \quad \text{in }[0,T]\times\R^n\times\R^n,
        \label{eq:property_ConvConj}
    \end{align}
    where 
    \begin{align}
        &(L^b)^{*} (s,x,p) \coloneqq \max_b [p\cdot b - L^b (s,x,b)],\notag\\
        &(L^b)^{**} (s,x,b) \coloneqq \max_p [p\cdot b - (L^b)^* (s,x,p)],\notag\\
        & H^* (s,x,b) \coloneqq \max_p [p\cdot b -H(s,x,p)],
        \label{eq:DEF_Hamil_ConvConj}
    \end{align}
    and
    \begin{align}
        \text{Dom}(H^* (s,x,\cdot)) = \text{Conv}(B(s,x)).
        \label{eq:Dom_ConvCtrl}
    \end{align}
    $(L^b)^* $ and $H^*$ are the Legendre-Fenchel transformations (convex conjugate) of $L^b$ and $H$, respectively, with respect to $p$ for each $(s,x)$.
    \text{Dom}$(H^* (s,x,\cdot))$ represents the domain of $H^* (s,x,\cdot)$, and \text{Conv}$(B(s,x))$ represents the convex hull of $B(s,x)$.
    \label{lemma:Lemma1}
\end{lemma}
\noindent
\textbf{Proof.}  See Appendix \ref{appen:Lemma1_proof}.

\eqref{eq:Dom_ConvCtrl} in Lemma \ref{lemma:Lemma1} implies that the domain of $H^*(s,x,\cdot)$ contains the domain of $L^b(s,x,\cdot)$ for each $(s,x)$.
Even though \text{Conv}$(B(s,x))$ is convex in $b\in\R^n$, $\text{Dom} (H^*(s,\cdot,\cdot)) = \{ (x,b)~|~ b\in \text{Conv}(B(s,x))  \}$ is generally non-convex in $(x,b)\in\R^n\times\R^n$ for each $s\in[t,T]$.

To utilize Lemma \ref{lemma:Lemma1}, we now propose the generalized Lax formula for the state-unconstrained problems (\eqref{eq:NoConst_vartheta_cost} subject to \eqref{eq:NoConst_vartheta_const}) in Theorem \ref{thm:GHopfLax}.

\begin{theorem} \textnormal{\textbf{(The generalized Lax formula for the state-unconstrained optimal control problem)}}
    Suppose Assumption \ref{assum:BigAssum} holds. For given initial time and state $(t,x)\in[0,T]\times\R^n$, define 
    \begin{align}
        \bar{\vartheta}(t,x):=\inf_{\beta} \int_t^T H^*(s,\mathrm{x}(s),\beta(s)) ds + g(\mathrm{x}(T)),
        \label{eq:HopfLax_cost}\\
        \text{subject to } 
        \begin{cases}
            \dot{\mathrm{x}}(s) = -\beta(s), & s\in [t,T], \\
            \mathrm{x}(t) = x,\\
            \beta(s) \in \text{Conv}(B(s,\mathrm{x}(s))), & s\in [t,T],\\
        \end{cases} 
        \label{eq:HopfLax_const}
    \end{align}
    where $H^*$ and $B$ are defined in \eqref{eq:DEF_Hamil_ConvConj} and \eqref{eq:Trans_CtrlSet}, respectively.
    Then,
    \begin{align}
        \vartheta(t,x) = \bar{\vartheta}(t,x) \quad \forall (t,x)\in[0,T]\times\mathbb{R}^n,
    \end{align}
    where $\vartheta$ is defined in \eqref{eq:NoConst_vartheta_cost} and \eqref{eq:NoConst_vartheta_const}.
    \label{thm:GHopfLax}
\end{theorem}
\textbf{Proof.} This proof generalizes the proof for the Hopf-Lax formula presented in \cite{bardi1984hopf}.
By the HJB PDE in Theorem \ref{thm:HJB_PDE}, $\bar{\vartheta}$ is the viscosity solution to
\begin{align*}
    \bar{\vartheta}_t (t,x) - \max_{b\in \text{Conv}(B(t,x))} [D\bar{\vartheta}(t,x)\cdot b - H^* (t,x,b)] =0
\end{align*}
in $(0,T)\times \R^n$, and $\bar{\vartheta}(T,x)=g(x)$ on $\{t=T\}\times\mathbb{R}^n$.
Since $H$ in \eqref{eq:Hamiltonian} is convex and semi lower-continuous in $p$, and \eqref{eq:Dom_ConvCtrl} in Lemma \ref{lemma:Lemma1} holds, then 
\begin{small}
\begin{align*}
    \max_{b\in \text{Conv}(B(t,x))} [D\bar{\vartheta}(t,x)\cdot b - H^* &(t,x,b)] 
    = H(t,x,D\bar{\vartheta}(t,x)).
\end{align*}
\end{small}

Theorem \ref{thm:HJB_PDE} states that $\vartheta$ is the unique viscosity solution to
\begin{align*}
    \vartheta_t (t,x) - H(t,x,D\vartheta(t,x)) =0
\end{align*}
in $\R^n\times (0,T)$, and $\bar{\vartheta}(T,x)=g(x)$ on $\{t=T\}\times\mathbb{R}^n$.
$\vartheta$ and $\bar{\vartheta}$ solve the same HJB PDE with the same terminal cost, hence, $\vartheta\equiv \bar{\vartheta}$ by the solution uniqueness.
$\qed$

\begin{remark}
    The state-unconstrained optimal control problem (\eqref{eq:NoConst_vartheta_cost} subject to \eqref{eq:NoConst_vartheta_const}) can be solved by the generalized Lax formula in Theorem \ref{thm:GHopfLax}.
    \label{remark:OptProblem_GHopfLax}
\end{remark}

\subsection{Optimal control analysis using the generalized Lax formula}
\label{sec:OptCtrl_NoStateConst}

In this subsection, we present in Theorem \ref{thm:OptCtrl_PostProcess} the relationship between a feasible state trajectory and a control signal  for the state-unconstrained problem ($\vartheta$ in \eqref{eq:NoConst_vartheta_cost} subject to \eqref{eq:NoConst_vartheta_const}) and those for the generalized Lax formula ($\bar{\vartheta}$ in Theorem \ref{thm:GHopfLax}).

We first state some properties regarding the Legendre-Fenchel transformations in Lemma \ref{lemma:lemma6}.

\begin{lemma} [Decomposition of control and stage cost] 
    Suppose Assumption \ref{assum:BigAssum} holds.
    For all $(s,x)\in[0,T]\times\R^n$ and $b\in \text{Conv}(B(s,x))$, there exist a finite $b_i \in B(s,x)$, $a_i \in A$, $\gamma_i\in\R$ such that
    \begin{align}
        &H^* (s,x,b) = \sum_i \gamma_i L^b ( s,x,b_i) = \sum_i \gamma_i L (s,x,a_i),\label{eq:LinearComb_cost}\\
        &\quad\quad\quad\quad\quad b = \sum_i \gamma_i b_i,\quad b_i = -f(s,x,a_i), \label{eq:LinearComb_cost2}
    \end{align}
    where $L^b (s,x,b_i ) = L(s,x,a_i)$, $\sum_i \gamma_i = 1$, and $\gamma_i \geq 0$.
    Note that $A$ is the control constraint in \eqref{eq:NoConst_vartheta_const}, and $B(s,x)$ is defined in \eqref{eq:Trans_CtrlSet}.
    \label{lemma:lemma6}
\end{lemma}
\noindent
\textbf{Proof.} The convex hull of $B(s,x)$ is identical to the set of all convex combinations of $B(s,x)$. 
Since Lemma \ref{lemma:Lemma1} holds and $B(s,x)$ is bounded for each $(s,x)\in[t,T]\times\R^n$, there exist a finite number of $b_i\in B(s,x)$ and $\gamma_i$ such that
\begin{align*}
    H^*(s,x,b) &= \sum_i \gamma_i L^b ( s,x,b_i),\quad
    b = \sum_i \gamma_i b_i,
\end{align*}
where $\sum_i \gamma_i = 1$ and $\gamma_i \geq 0$. By \eqref{eq:Def_Lv}, for each $i$, there exists $a_i\in A$ such that $L^b (s,x,b_i) = L(s,x,a_i )$. $\qed$

Consider any feasible control signal ($\beta$) and state ($\mathrm{x}$) trajectories solving \eqref{eq:HopfLax_const}.
Corresponding to $\beta$ and $\mathrm{x}$, Theorem \ref{thm:OptCtrl_PostProcess} proposes a corresponding control signal ($\alpha^\epsilon\in\mathcal{A}(t)$) and approximate state trajectory ($\mathrm{x}^\epsilon$) solving \eqref{eq:NoConst_vartheta_const} such that $\|\mathrm{x}-\mathrm{x}^\epsilon\|_{L^\infty(t,T)}<\epsilon$, and the difference of the costs in \eqref{eq:HopfLax_cost} and \eqref{eq:NoConst_vartheta_cost} is
\begin{align}
    &\bigg| \int_t^T H^*(s,\mathrm{x}(s),\beta(s)) ds + g(\mathrm{x}(T)) \notag\\
    &\quad\quad\quad\quad -\int_t^T L(s,\mathrm{x}^\epsilon(s),\alpha^\epsilon(s)) ds  -g(\mathrm{x}^\epsilon(T))\bigg| <\epsilon.
    \label{eq:cost_err}
\end{align}

Assume that $\beta$ is Riemann integrable in $(t,T)$. For some $\delta>0$, consider a temporal discretization: $\{t_0=t,...t_K=T\}$ such that $\Delta t_k\coloneqq t_{k+1}-t_k <\delta,\forall k=0,...,K-1$.
We define a control signal $\alpha^\epsilon\in \mathcal{A}(t)$: for $k=\{0,...,K-1\}$,
\begin{align}
\begin{small}
    \alpha^\epsilon(s) = a_i^k,\quad s\in\bigg[t_k + \sum_{j=1}^{i-1}\gamma_j^k \Delta t_k,t_k + \sum_{j=1}^{i}\gamma_j^k \Delta t_k \bigg),
\end{small}
\label{eq:approx_ctrl}
\end{align}
where $a_i^k$ and $\gamma_i^k$ are $i$-th control and coefficient in Lemma \ref{lemma:lemma6} for $t=t_k$, $x=\mathrm{x}(t_k)$, and $b=\beta(t_k)$. 
We also define a state trajectory $\mathrm{x}^\epsilon:[t,T]\rightarrow\R^n$ solving 
\begin{align}
    \dot{\mathrm{x}}^\epsilon(s) = f(s,\mathrm{x}^\epsilon(s),\alpha^\epsilon(s)), \quad s\in(t,T),\quad \mathrm{x}^\epsilon(t)=x.
    \label{eq:approx_traj}
\end{align}
Theorem \ref{thm:OptCtrl_PostProcess} states that $\alpha^\epsilon$ in \eqref{eq:approx_ctrl} and $\mathrm{x}^\epsilon$ in \eqref{eq:approx_traj} are control signal and approximate state trajectory that satisfy $\|\mathrm{x}-\mathrm{x}^\epsilon\|_{L^\infty(t,T)}<\epsilon$ and \eqref{eq:cost_err} for some small $\delta>0$.

\begin{theorem}
    Suppose Assumption \ref{assum:BigAssum} holds. For initial time and state $(t,x)\in[0,T]\times\R^n$, consider any feasible control signal $\beta$ and state trajectory $\mathrm{x}$ solving \eqref{eq:HopfLax_const}. Assume that $\beta$ is Riemann integrable in $[t,T]$.
    Then, for any $\epsilon>0$, there exists $\delta>0$ such that, for any discretization $\{t_0=t,...,t_K=T\}$ where $\lvert \Delta t_{k} \rvert < \delta,$ $k=0,...,K-1$: $\alpha^\epsilon$ in \eqref{eq:approx_ctrl} and $\mathrm{x}^\epsilon$ in \eqref{eq:approx_traj} satisfy
    \begin{align}
        \| \mathrm{x} - \mathrm{x}^\epsilon \|_{L^\infty (t,T)} < \epsilon 
        \label{eq:OptTraj_approximation}
    \end{align}
    and
    \begin{align}
        &\bigg| \int_t^T H^*(s,\mathrm{x}(s),\beta(s)) ds + g(\mathrm{x}(T)) \notag\\
        &\quad\quad\quad\quad -\int_t^T L(s,\mathrm{x}^\epsilon(s),\alpha^\epsilon(s)) ds  -g(\mathrm{x}^\epsilon(T))\bigg|<\epsilon.
        \label{eq:OptCtrl_approximation}
    \end{align}
    \label{thm:OptCtrl_PostProcess}
\end{theorem}
\noindent
\textbf{Proof.} See Appendix \ref{appen:Theorem_OptCtrl}.

\begin{remark}[Optimal control]
    ~
    \begin{enumerate}
        \item Given an optimal control signal for the generalized Lax formula,
        Theorem \ref{thm:OptCtrl_PostProcess} provides an approximate optimal control signal for the state-unconstrained problem (\eqref{eq:NoConst_vartheta_cost} subject to \eqref{eq:NoConst_vartheta_const}) with $\epsilon$-error bounds on the state trajectory and cost.
        
        \item As $\epsilon$ goes to 0, the errors on the state trajectory and cost in \eqref{eq:OptTraj_approximation} and \eqref{eq:OptCtrl_approximation}, respectively, converge to 0. However, the limit point of $\alpha^\epsilon$ might not exist.
    \end{enumerate}
    \label{remark:Exist_Approx_OptCtrl}
\end{remark}
Note that the existence of the optimal control signal $\alpha_*$ is not obvious, which is why the definition of the state-constrained problem \eqref{eq:NoConst_vartheta_cost} subject to \eqref{eq:NoConst_vartheta_const} uses the infimum instead of the minimum.
If $A$ is convex, there exists a minimizer $\alpha_*$ \cite{altarovici2013general}, otherwise, a minimizer might not exist.

\subsection{The generalized Lax formula for state-constrained problems}
\label{sec:NewFormulation_StateConst}

In this section, we extend the generalized Lax formula in Theorem \ref{thm:GHopfLax} for the state-constrained problem \eqref{eq:def_vartheta}.
For derivation, we utilize the theory and properties presented in Section \ref{sec:NewFormulation1} and \ref{sec:OptCtrl_NoStateConst}.
It will be also shown that Theorem \ref{thm:OptCtrl_PostProcess} is also valid for the state-constrained problem \eqref{eq:def_vartheta}.

Theorem \ref{thm:Value_Equal_StateConstraint} presents the generalized Lax formula for the state-constrained optimal control problems in \eqref{eq:def_vartheta}.
\begin{theorem}\textnormal{\textbf{(The generalized Lax formula for the state-constrained optimal control problem)}}
    Suppose Assumption \ref{assum:BigAssum} holds. For initial time and state $(t,x)\in[0,T]\times\R^n$, define 
    \begin{align}
        \bar{\vartheta}(t,x) \coloneqq \inf_{\beta} \int_t^T  H^*(s,\mathrm{x}(s),\beta(s))ds + g(\mathrm{x}(T)),
        \label{eq:StateConst_Trans_ValueFunc}\\
        \text{subject to } 
        \begin{cases}
            \dot{\mathrm{x}}(s) = -\beta(s), &s\in [t,T], \\
            \mathrm{x}(t) = x,\\
            \beta(s)\in \text{Conv}(B(s,\mathrm{x}(s))),  &s\in [t,T],\\
            c(s,\mathrm{x}(s)) \leq 0, &s\in [t,T].
        \end{cases} 
        \label{eq:StateConst_Trans_Constraint}    
    \end{align}
    Then,
    \begin{align}
        \vartheta(t,x) = \bar{\vartheta}(t,x)\quad \text{in }[t,T]\times\R^n.
    \end{align}
    \label{thm:Value_Equal_StateConstraint}
\end{theorem}

The proof of Theorem \ref{thm:Value_Equal_StateConstraint} requires some development.
Corresponding to $\vartheta$ in \eqref{eq:def_vartheta}, we define functions $J$ and $V$ with the auxiliary variable $z\in\R$: for initial time $t\in[0,T]$, initial state $x\in\R^n$, auxiliary variable $z\in\R$, control signal $\alpha\in\mathcal{A}(t)$,
\begin{align}
    & J(t,x,z,\alpha)  \coloneqq \max\bigg\{ \max_{s\in[t,T]} c(s,\mathrm{x}(s)) ,\notag \\& \quad\quad\quad \int_t^T L(s,\mathrm{x}(s),\alpha(s))ds+g(\mathrm{x}(T))-z \bigg\},
    \label{eq:Def_J}
\end{align}
where $\mathrm{x}$ and $\alpha$ solves \eqref{eq:NoConst_vartheta_const}, and
\begin{align}
    V(t,x,z)  \coloneqq \inf_{\alpha\in\mathcal{A}(t)}J(t,x,z,\alpha).
    \label{eq:Def_V}
\end{align}
In $J$ in \eqref{eq:Def_J}, the cost \eqref{eq:def_vartheta_ep} and the state constraint \eqref{eq:def_vartheta_ep_const} of $\vartheta$ are combined together into a single cost.
We also define a function $\bar{J}$ and $\bar{V}$: for initial time $t\in[0,T]$, initial state $x\in\R^n$, auxiliary variable $z\in\R$, control signal $\beta\in\{\beta ~|~\beta(s)\in \text{Conv}(B(s,\mathrm{x}(s)))\}$, where $\mathrm{x}$ and $\beta$ solves \eqref{eq:HopfLax_const}, 
\begin{align}
    & \bar{J}(t,x,z,\beta)  \coloneqq \max\bigg\{ \max_{s\in[t,T]} c(s,\mathrm{x}(s)) ,\notag \\& \quad\quad\quad \int_t^T H^*(s,\mathrm{x}(s),\beta(s))ds+g(\mathrm{x}(T))-z \bigg\},
    \label{eq:Def_BarJ}
\end{align}
and
\begin{align}
    \bar{V}(t,x,z)  \coloneqq \inf_{\beta}\bar{J}(t,x,z,\beta)
    \label{eq:Def_BarV}
\end{align}
subject to $\beta\in\{\beta ~|~\beta(s)\in \text{Conv}(B(s,\mathrm{x}(s)))\}$.
In $\bar{J}$ in \eqref{eq:Def_BarJ}, the cost \eqref{eq:StateConst_Trans_ValueFunc} and the state constraint \eqref{eq:StateConst_Trans_Constraint} of $\bar{\vartheta}$ are also combined.
Then, $\vartheta$ in \eqref{eq:def_vartheta}, $V$ in  \eqref{eq:Def_V}, $\bar{\vartheta}$ in \eqref{eq:StateConst_Trans_ValueFunc} subject to \eqref{eq:StateConst_Trans_Constraint}, and $\bar{V}$ in \eqref{eq:Def_BarV} satisfy the following properties.

\begin{lemma} Suppose Assumption \ref{assum:BigAssum} holds. For given $(t,x)\in[0,T]\times\R^n$, 
\begin{align}
    & \vartheta(t,x) = \min z \text{  subject to  } V(t,x,z)\leq 0,\label{eq:theta_equiv}\\
    & \bar{\vartheta}(t,x) = \min z \text{  subject to  } \bar{V}(t,x,z)\leq 0.
    \label{eq:bartheta_equiv}
\end{align}
where $\vartheta$, $\bar{\vartheta}$, $V$, $\bar{V}$ are defined in \eqref{eq:def_vartheta}, \eqref{eq:StateConst_Trans_ValueFunc}, \eqref{eq:Def_V}, and \eqref{eq:Def_BarV}, respectively.
\label{lemma:augmentedValueFunc}
\end{lemma}
\textbf{Proof.} The proof for \eqref{eq:bartheta_equiv} can be found in \cite{altarovici2013general} since $\text{Conv}(B(s,\mathrm{x}(s)))$ is convex in $b$.
We will prove \eqref{eq:theta_equiv} by showing that $\vartheta(t,x)-z \leq 0 \Leftrightarrow V(t,x,z)\leq 0$.

\noindent (i) $\vartheta(t,x)-z\leq 0 \Rightarrow V(t,x,z)\leq 0$ 

For $(t,x,z)$ satisfying $\vartheta(t,x)-z\leq 0$, $\vartheta(t,x)$ is finite.
Since $\vartheta^\epsilon$ is increasing as $\epsilon$ goes to 0, 
\begin{align}
    \vartheta^{\epsilon}(t,x) \leq \vartheta(t,x) < z.
    \label{eq:thm1_proof1}
\end{align}
For any small $\epsilon_1>0$, there exists a feasible $\alpha^{\epsilon_1}\in\mathcal{A}(t)$ such that 
\begin{align}
    \int_t^T L(s,\mathrm{x}^{\epsilon_1}(s),\alpha^{\epsilon_1}(s))ds +g(\mathrm{x}^{\epsilon_1}(T)) \leq \vartheta^{\epsilon} (t,x) +\epsilon_1
    \label{eq:thm1_proof2}
\end{align}
and 
\begin{align}
    \max_{s\in[t,T]}c(s,\mathrm{x}^{\epsilon_1}(s)) \leq \epsilon,
    \label{eq:thm1_proof3}
\end{align}
where $\mathrm{x}^{\epsilon_1}$ solves the dynamics in \eqref{eq:def_vartheta_ep_const} with $\alpha^{\epsilon_1}$.
Note that $\mathcal{A}(t)$ is defined in \eqref{eq:def_CtrlTraj}. 

By \eqref{eq:thm1_proof1}, \eqref{eq:thm1_proof2}, and \eqref{eq:thm1_proof3}, we have
\begin{align*}
    \max\{ \epsilon , \epsilon_1 &\} \geq \max\bigg\{ \max_{s\in[t,T]}c(s,\mathrm{x}^{\epsilon_1}(s)), \\
    &\int_t^T L(s,\mathrm{x}^{\epsilon_1}(s),\alpha^{\epsilon_1}(s))ds + g(\mathrm{x}^{\epsilon_1}(T))-z \bigg\}
\end{align*}
for any $\epsilon_1>0$. As $\epsilon, \epsilon_1 \rightarrow0$, $V(t,x,z)\leq 0$.
\\\\
\noindent (ii) $V(t,x,z)\leq 0 \Rightarrow \vartheta(t,x)-z\leq 0$ 

For any $\epsilon>0$, there exists $\alpha^\epsilon\in\mathcal{A}(t)$ such that
\begin{align*}
    \epsilon \geq \max\bigg\{ &\max_{s\in[t,T]}c(s,\mathrm{x}^\epsilon(s)), \\
    &\int_t^T L(s,\mathrm{x}^\epsilon(s),\alpha^\epsilon(s))ds + g(\mathrm{x}^\epsilon(T))-z \bigg\},
\end{align*}
where $\mathrm{x}^\epsilon$ solves the dynamics in \eqref{eq:def_vartheta_ep_const} with $\alpha^\epsilon$. This implies that $\vartheta^\epsilon(t,x) \leq z+\epsilon$. As $\epsilon$ goes to 0, we have $\vartheta(t,x)-z\leq 0$.
$\qed$

We are ready to prove Theorem \ref{thm:Value_Equal_StateConstraint}.
\\
\noindent \textbf{Proof.} We will prove that $V(t,x,z)=\bar{V}(t,x,z)$, then, by Lemma \ref{lemma:augmentedValueFunc}, Theorem \ref{thm:Value_Equal_StateConstraint} is proved.

\noindent
(i) $V(t,x,z) \geq \bar{V}(t,x,z)$

For any feasible state ($\mathrm{x}$) and control ($\alpha$) trajectories solving \eqref{eq:NoConst_vartheta_const}, define a control signal ($\beta$): 
\begin{align*}
    \beta(s) = -f(s,\mathrm{x}(s),\alpha(s))\in B(s,\mathrm{x}(s)),\quad s\in[t,T].
\end{align*}
Then, $\mathrm{x}$ and $\beta$ solve $\dot{\mathrm{x}}(s)=-\beta(s)$ for $s\in[t,T]$ and $\mathrm{x}(t)=x$. 
By Lemma \ref{lemma:Lemma1}, $L(s,\mathrm{x}(s),\alpha(s))=L^b(s,\mathrm{x}(s),\beta(s))\geq H^*(s,\mathrm{x}(s),\beta(s))$ for all $s\in[t,T]$. This implies that
\begin{align*}
    J(t,x,z,\alpha) \geq \bar{J}(t,x,z,\beta) \geq \bar{V}(t,x,z).
\end{align*}
Since the above inequality holds for all any feasible $\mathrm{x}$ and $\alpha$, we conclude $V(t,x,z) \geq \bar{V}(t,x,z)$.

\noindent (ii) $V(t,x,z)\leq \bar{V}(t,x,z)$

For any feasible state ($\mathrm{x}$) and control ($\beta$) trajectories solving \eqref{eq:HopfLax_const}, by Theorem \ref{thm:OptCtrl_PostProcess}, there exists $\mathrm{x}^\epsilon$ and $\alpha^\epsilon$ solving \eqref{eq:NoConst_vartheta_const} such that \eqref{eq:approx_traj} and \eqref{eq:approx_ctrl} hold for any $\epsilon>0$. Then,
\begin{align}
    \bar{J}(t,x,z,\beta) &\geq J(t,x,z,\alpha^\epsilon)-\max\{1,L_c\}\epsilon\notag\\
    & \geq V(t,x,z) - \max\{1,L_c\}\epsilon,
    \label{eq:proof_Thm3_1}
\end{align}
where $L_c$ is the Lipschitz constant for $c$ in Assumption \ref{assum:BigAssum}. 
Since \eqref{eq:proof_Thm3_1} holds for any $\mathrm{x}$, $\beta$ solving \eqref{eq:HopfLax_const} and $\epsilon>0$, $\bar{V}(t,x,z)\geq V(t,x,z)$.
$\qed$
\\
\noindent
In Theorem \ref{thm:Value_Equal_StateConstraint}, we observe that the state constraint in the state-constrained optimal control problem \eqref{eq:def_vartheta} and the generalized Lax formula (\eqref{eq:StateConst_Trans_ValueFunc} subject to \eqref{eq:StateConst_Trans_Constraint}) are the same.


\begin{remark}
~
\begin{enumerate}
    \item The state-constrained optimal control problem \eqref{eq:def_vartheta} can be solved by the generalized Lax formula in Theorem \ref{thm:Value_Equal_StateConstraint}.
    
    \item Given an optimal state trajectory and control signal for the generalized Lax formula in Theorem \ref{thm:Value_Equal_StateConstraint}, Theorem \ref{thm:OptCtrl_PostProcess} provides approximate optimal state trajectory and control signal for the state-constrained problem \eqref{eq:def_vartheta} such that the approximation error of the cost is less than $\epsilon$.
\end{enumerate}
\label{remark:OptCtrl_StateConst}
\end{remark}
The second point of Remark \ref{remark:OptCtrl_StateConst} implies that the optimal control analysis for the state-unconstrained problems (\eqref{eq:NoConst_vartheta_cost} subject to \eqref{eq:NoConst_vartheta_const}) in Section \ref{sec:OptCtrl_NoStateConst} is valid for the state-constrained problems \eqref{eq:def_vartheta}, which is found in the proof of Theorem \ref{thm:Value_Equal_StateConstraint}.

\subsection{Numerical Algorithm}

Algorithm \ref{alg:Opt_NewFormulation1} presents a numerical algorithm to compute an optimal state trajectory ($\mathrm{x}$) and a control signal ($\alpha$) for the state-constrained problem \eqref{eq:def_vartheta} using the generalized Lax formula in Theorem \ref{thm:Value_Equal_StateConstraint}. 

\begin{algorithm}
\caption{Computing optimal state trajectory ($\mathrm{x}$) and control signal ($\alpha$) for the state-constrained problem \eqref{eq:def_vartheta} using the generalized Lax formula}
\begin{algorithmic}[1]
\State \textbf{Input:} {initial time $t$, initial state $x$}
\State \textbf{Output:} {the optimal state ($\mathrm{x}$) and control ($\alpha$) trajectories}
\State Generate a temporal discretization: $\{t_0=t,...,t_K=T\}$
\State Solve \eqref{eq:Numerical_HopfLax_Cost} subject to \eqref{eq:Numerical_HopfLax_Const}, and get $\mathrm{x}_* [\cdot],\beta_*[\cdot]$ 
\State Find $(a_i^k,b_i^k,\gamma_i^k)$ solving \eqref{eq:LinearComb_cost} and \eqref{eq:LinearComb_cost2} for $x=\mathrm{x}_*[k]$ and $s=t_k$
\State Additionally discretize each temporal interval: \begin{small}\begin{align}
    [t_k,t_{k+1})=\bigcup_i [t_k+\sum_{j=1}^{i-1}\gamma_j^k \Delta t_k, t_k+\sum_{j=1}^{i}\gamma_j^k \Delta t_k)
\end{align}\end{small}
\State Design $\alpha_*^\epsilon$ using $a_i^k$ by \eqref{eq:approx_ctrl} and compute $\mathrm{x}_*^\epsilon$ by solving the ODE \eqref{eq:approx_traj}
\end{algorithmic}
\label{alg:Opt_NewFormulation1}
\end{algorithm}

We first numerically compute an optimal state trajectory ($\mathrm{x}$) and a control signal ($\beta$) for the generalized Lax formula in Theorem \ref{thm:Value_Equal_StateConstraint}, and then utilize Theorem \ref{thm:OptCtrl_PostProcess} to get a numerical optimal state ($\mathrm{x}^\epsilon$) and control ($\alpha^\epsilon$) for the state-constrained problem \eqref{eq:def_vartheta}.

Numerical optimization methods, such as the interior-point method \cite{boyd2004convex}, can be utilized to compute an optimal state trajectory ($\mathrm{x}$) and a control signal ($\beta$) for the generalized Lax formula using the temporal discretization $\{t_0=t,...,t_K =T\}$:
\begin{align}
    &\bar{\vartheta}(t,x) \simeq \min_{\mathrm{x}[\cdot],\beta[\cdot]} \sum_{k=0}^{K-1} H^*(t_k,\mathrm{x}[k],\beta[k])\Delta_k + g(\mathrm{x}[K]),\label{eq:Numerical_HopfLax_Cost}\\
    &\text{subject to }\begin{cases}\mathrm{x}[k+1]=\mathrm{x}[k]-\beta[k]\Delta_k,&k=0,...,K-1,\\\mathrm{x}[0]=x,\\\beta[k]\in\text{Conv}(B(t_k,\mathrm{x}[k])),&k=0,...,K-1,\\c(t_k,\mathrm{x}[k])\leq 0,&k=0,...,K-1. \end{cases}
    \label{eq:Numerical_HopfLax_Const}
\end{align}
This optimization problem is defined in the sequence of states and controls: $\mathrm{x}[\cdot],\beta[\cdot]$. 
If \eqref{eq:Numerical_HopfLax_Cost} subject to \eqref{eq:Numerical_HopfLax_Const} is convex, gradient-based methods efficiently provide the global optimality without discretization in state space.

For optimal state and control sequences of the problem (\eqref{eq:Numerical_HopfLax_Cost} subject to \eqref{eq:Numerical_HopfLax_Const}), we denote $\mathrm{x}_*[\cdot],\beta_*[\cdot]$.
Using Lemma \ref{lemma:lemma6} and Theorem \ref{thm:OptCtrl_PostProcess}, we find $(a_i^k,b_i^k,\gamma_i^k)$ solving \eqref{eq:LinearComb_cost} and \eqref{eq:LinearComb_cost2} for $x=\mathrm{x}_*[k],b=\beta_*[k]$, and design an approximate optimal control signal ($\alpha_*^\epsilon$) by \eqref{eq:approx_ctrl} using $a_i^k$.
Then, we get the corresponding state trajectory $\mathrm{x}^\epsilon_*$ by solving \eqref{eq:approx_traj} for $\alpha_*^\epsilon$.
Note that, to get $\alpha_*^\epsilon$ using \eqref{eq:approx_ctrl}, an additional temporal discretization $\{t_k,...,t_k+\sum_{j=1}^{i}\gamma_j^k\Delta t_k,...,t_{k+1}\}$ is necessary in each time interval $[t_k,t_{k+1})$, $k=0,...,K-1$.

This additional discretization causes a frequent control switching in short time, which might cause some practical issues.
It could be possible to reduce the control switching by removing the process of the additional discretization: find $\alpha_*[\cdot]$ over the discretization $\{t_0=t,...,t_K=T\}$. 
Although we do not have theoretical proof, we provide some practical suggestions as follows: 1) using $(a_i^k,b_i^k,\gamma_i^k)$ solving \eqref{eq:LinearComb_cost} and \eqref{eq:LinearComb_cost2}, pick a maximum likelihood control $a_{i_*}^k$ where $i_* =\arg\max_i \gamma_i^k$, i.e., $\alpha_* [k] = a_{i_*}^k$;
2) if the stage cost $L$ does not depend on the control, find
\begin{align}
    \alpha_*[k] \in\arg\min_{a\in A} \| f(t_k,\mathrm{x}_*^\epsilon[k],a) \Delta_k - \mathrm{x}_*[k+1]  + \mathrm{x}_*^\epsilon[k]\|, 
\end{align}
where $\mathrm{x}_*[k+1]=\mathrm{x}_*[k]-\beta_*[k]\Delta t_k$, $\mathrm{x}_*^\epsilon[k+1]=\mathrm{x}_*^\epsilon[k] + f(t_k,\mathrm{x}_*^\epsilon [k],\alpha_*[k])\Delta t_k$.
This is equivalent to finding a control that minimizes $\|\mathrm{x}^\epsilon_*[k+1]-\mathrm{x}_*[k+1]\|$.
These methods filter out the control switching and find a feasible control as well. 
One topic of our current work is to theoretically validate our suggestions or to propose other methods for the control switching issue.

\section{Convexity Analysis for the generalized Lax Formula}
\label{sec:CvxAnalysis_GeneralizedHopfLax}

The generalized Lax formula with the numerical method in Section \ref{sec:NewFormulation} allows efficient computation if the temporally discretized generalized Lax formula (\eqref{eq:Numerical_HopfLax_Cost} subject to \eqref{eq:Numerical_HopfLax_Const}) is convex.
Thus, in this section, we analyze the convexity conditions of this problem.
In this convexity analysis, we present the benefits of solving the generalized Lax formula in comparison to solving the state-constrained problem \eqref{eq:def_vartheta} using the temporal discretization:
\begin{align}
    &\quad \vartheta(t,x) \simeq \min_{\mathrm{x}[\cdot],\alpha[\cdot]} \sum_{k=0}^{K-1} L(k,\mathrm{x}[k])\Delta_k + g(\mathrm{x}[K]),\label{eq:Numerical_Problem_Cost}\\
    &\text{subject to }\begin{cases}\mathrm{x}[k+1]=\mathrm{x}[k]+f(t_k,\mathrm{x}[k],\alpha[k])\Delta_k,\\\mathrm{x}[0]=x,\\\alpha[k]\in A,\\c(t_k,\mathrm{x}[k])\leq 0, \end{cases}
    \label{eq:Numerical_Problem_Const}
\end{align}
where $\mathrm{x}[\cdot],\alpha[\cdot]$ are the state and control sequences.
\subsection{Convexity analysis for the generalized Lax formula}

For convexity analysis, we deal with the systems whose stage cost is in the following form:
\begin{equation}
    L(s,x,a)=L^x(s,x)+L^a(s,a),
    \label{eq:Stagecost_Decom}
\end{equation}
where $L$ is the stage cost of the state-constrained optimal control problem \eqref{eq:def_vartheta}.
The convexity conditions for the temporally discretized state-constrained problem (\eqref{eq:Numerical_Problem_Cost} subject to \eqref{eq:Numerical_Problem_Const}) are given in Condition \ref{cvxCondition:ConvexCondition_original}.

\begin{cvxCondition} \textnormal{\textbf{(The convexity conditions for the temporally discretized state-constrained problem)}}
    Suppose \eqref{eq:Stagecost_Decom} holds. 
    $L$, $g$, $A$, $f$, and $c$ are the stage cost, terminal cost, control constraint, dynamics, and state constraint, respectively, for the state-constrained problem \eqref{eq:def_vartheta}.
    \begin{enumerate}
        \item  For each $s\in[t,T]$, $L^x(s,\cdot)$ is convex in $x\in\R^n$ and $L^a(s,\cdot)$ is convex in $a\in A\subset\R^m$,
        
        \item $g$  is convex in  $x \in \R^n$,
        
        \item $A$ is convex,
        
        \item $f(s,\cdot,\cdot)$ is affine in $(x,a) \in \R^n \times A$ for each $s\in[t,T]$,
        
        \item $c(s,\cdot)$ is convex in $x\in \R^n$ for each $s\in[t,T]$.
        
    \end{enumerate}
    \label{cvxCondition:ConvexCondition_original}
\end{cvxCondition}

In this subsection, we present a sufficient convexity condition for the temporally discretized generalized Lax formula (\eqref{eq:Numerical_HopfLax_Cost} subject to \eqref{eq:Numerical_HopfLax_Const}) with respect to the stage cost, terminal cost, dynamical constraint, control constraint, and state constraint in Table \ref{tab:class2}. 

We present Lemma \ref{lemma:NewStagecost_Decomp} that will be used to derive a convexity condition of the stage cost for the temporally discretized generalized Lax formula (\eqref{eq:Numerical_HopfLax_Cost} subject to \eqref{eq:Numerical_HopfLax_Const}) in Lemma \ref{lemma:StageCost_cvxCondition}.
\begin{lemma}
    Suppose \eqref{eq:Stagecost_Decom} holds. Then,
    \begin{align}
        H^*(s,x,b) = L^x (s,x) + (H^a)^* (s,x,b),
        \label{eq:HamConj_Decomp}
    \end{align}
    where $H^*$ is defined in \eqref{eq:DEF_Hamil_ConvConj},
    \begin{align}
        &H^a(s,x,p) \coloneqq \max_{a\in A} -p\cdot f(s,x,a) - L^a (s,a),\label{eq:HamConj_Decomp_2}\\
        &(H^a)^*(s,x,b) \coloneqq \max_{p} p\cdot b - H^a(s,x,p).
        \label{eq:HamConj_Decomp_3}
    \end{align}
    \label{lemma:NewStagecost_Decomp}
\end{lemma}
\textbf{Proof. } By \eqref{eq:Def_Lv}, 
\begin{align*}
    L^b(s,x,b) = L^x (s,x) + (L^a)^b (s,x,b), 
\end{align*}
where $(L^a)^b (s,x,b)=\min_a L^a(s,a)$ subject to $ f(s,x,a) = -b $. 
By $(L^b)^* \equiv H$ by Lemma \ref{lemma:Lemma1} and the definition of $H$ in \eqref{eq:Hamiltonian},
\begin{align*}
    H(s,x,p) & = \max_b p\cdot b -  L^x (s,x) - (L^a)^b (s,x,b)\\
             & = -L^x (s,x) + H^a (s,x,p),
\end{align*}
where $H^a(s,x,p) = ((L^a)^b)^* (s,x,p)$. Then, 
\begin{align*}
    H^*(s,x,b)& = \max_p b\cdot p + L^x (s,x) - H^a (s,x,p) \notag\\
    & = L^x (s,x) + (H^a)^* (s,x,b).\qedd
\end{align*}
This shows that the stage cost of the generalized Lax formula ($H^*$) is decomposed into the control-independent ($L^x$) and control-dependent ($(H^a)^*$) parts similar to the stage cost of the state-constrained problem \eqref{eq:def_vartheta} ($L$) as in \eqref{eq:Stagecost_Decom}. 
One observation here is that the control-independent stage cost of the generalized Lax formula ($L^x$) is exactly the same as that of the state-constrained problem ($L^x$). 
\begin{corollary}
    If $L(s,x,a)=L^x (s,x)$,
    \begin{align}
        H^*(s,x,b)= L (s,x).
    \end{align}
    \label{corollary:NewStagecost_NoControlInput}
\end{corollary}
\textbf{Proof.} In the proof of Lemma \ref{lemma:NewStagecost_Decomp}, we need to set $L^a \equiv 0$ and $(L^a)^b \equiv 0$. Then, $(H^a)^*(s,x,b)=((L^a)^b)^{**}(s,x,b)=0$.$\qed$

Lemma \ref{lemma:StageCost_cvxCondition} presents the convexity condition for the stage cost of the generalized Lax formula.
\begin{lemma} [Convexity of the stage cost]
    Suppose Assumption \ref{assum:BigAssum} and \eqref{eq:Stagecost_Decom} hold.
    If $L^x (s,\cdot)$ is convex in $x\in\R^n$ for each $s\in[t,T]$ and the dynamics is in the following form:
    \begin{equation}
        f(s,x,a) = M(s)x + \varphi (s,a),
        \label{eq:Dynamics_Decom}
    \end{equation}
    where $f$ is the dynamics in \eqref{eq:def_vartheta_ep_const} and $M(\cdot)$ is a time-varying linear matrix,
    then
    \begin{align}
        H^*(s,x,b) = L^x (s,x) + (\bar{H}^a)^* (s,b+M(s)x),
    \end{align}
    where 
    \begin{align}
        &\bar{H}^a (s,p) \coloneqq \max_{a\in A} [-p\cdot \varphi(s,a) - L^a (s,a) ],\\
        &(\bar{H}^a)^* (s,b) \coloneqq \max_{p} [p\cdot b -\bar{H}^a  (s,p) ],
    \end{align}
    and 
    $H^* (s,\cdot,\cdot)$ is convex in $(x,b)$ for each $s\in[t,T]$. 
    \label{lemma:StageCost_cvxCondition}
\end{lemma}
Note that $\bar{H}^a (s,p)$ is independent on $x$ and convex in $p$ for each $t$. 
\\
\noindent
\textbf{Proof.}
By \eqref{eq:HamConj_Decomp_2} and \eqref{eq:HamConj_Decomp_3},
\begin{align*}
    &H^a (s,x,p) = -p\cdot (M(s)x) + \bar{H}^a (s,p),\\
    &(H^a)^* (s,x,b) = \max_{b} p\cdot (b+M(s)x) - \bar{H}^a (s,p)\\
    &\quad\quad\quad\quad\quad~~  = (\bar{H}^a)^* (s,b+M(s)x).
\end{align*}
By Lemma \ref{lemma:NewStagecost_Decomp}, 
\begin{align*}
    H^*(s,x,b) = L^x(s,x)+(\bar{H}^a)^* (s,b+M(s)x).
\end{align*}
Since $(\bar{H}^a)^*(s, \cdot)$ is convex in $b$ and $b+M(s)x$ is affine in $(x,b)$, $(H^a)^* (s,\cdot,\cdot)$ is convex in $(x,b)$. 
Therefore, $H^* (s,\cdot,\cdot)$ is convex in $(x,b)$ for each $s\in[t,T]$ if $L^x (s,\cdot)$ is convex in $x$ for each $s\in[t,T]$.
$\qed$

We define the control constraint of the generalized Lax formula in $(x,b)$-space: for $s\in[t,T]$,
\begin{align}
    \bar{B}(s) \coloneqq \{(x,b) ~|~ b\in \text{Conv}(B(s,x))\},
    \label{eq:Sum_V_set}
\end{align}
where $B(s,x)$ is defined in \eqref{eq:Trans_CtrlSet}.
Lemma \ref{lemma:CtrlSet_cvxCondition} presents convexity conditions for $\bar{B}(s)$.
    
\begin{lemma} [Convexity of the control constraint]
    Suppose Assumption \ref{assum:BigAssum} and \eqref{eq:Dynamics_Decom} holds.
    Then $\bar{B}(s)$ in \eqref{eq:Sum_V_set} is convex in $(x,b)$ for each $s\in[t,T]$.
    \label{lemma:CtrlSet_cvxCondition}
\end{lemma}
\noindent
\textbf{Proof.} We need to prove that, for $(x_1 , b_1)$, $(x_2, b_2)\in\bar{B}(s)$ and $d\in [0,1]$, 
\begin{align*}
     d b_1 + (1-d) b_2  \in \text{Conv}(B(s, dx_1 + (1-d)x_2)).
\end{align*}
This is equivalent to 
\begin{align*}
    d (b_1 + M(s) x_1) +(1-d)(b_2 &+ M(s)x_2) \\
    &\in \text{Conv}(\{-\varphi(s,a)~|~a\in A \}).
\end{align*}

Since $b_i \in \text{Conv}(B(s,x_i))$ for $i=1,2$, there exist a finite number of $a_{ij}$ and $\gamma_{ij} \in[0,1]$ ($\sum_j \gamma_{ij} =1$ for each $i$) such that
\begin{align*}
    b_i = -M(s) x_i - \sum_j \gamma_{ij} \varphi(s,a_{ij})
\end{align*}
for each $i=1,2$.
Using this, we have
\begin{align*}
    d b_1 + (1-d) b_2 &=- M(s)(d x_1 + (1-d)x_2) - \\
    &\sum_{j} [d \gamma_{1,j} \varphi(s, a_{1,j}) + (1-d) \gamma_{2,j}\varphi(s, a_{2,j})].
\end{align*}
Since $\text{Conv}(\{-\varphi(s,a)~|~a\in A \}$ is a convex set,
\begin{align*}
    d (b_1 + M(s) x_1) +(1-&d)(b_2 + M(s)x_2) \\
    &\in \text{Conv}(\{-\varphi(s,a)~|~a\in A \}). \qedd
\end{align*}

Remark \ref{remark:summary_convexity_GHL} summarizes the convexity conditions for the generalized Lax formula.
\begin{remark}\textnormal{\textbf{(Convexity of the generalized Lax formula)}}
    ~
    Suppose \eqref{eq:Stagecost_Decom} holds. 
    In the generalized Lax formula (\eqref{eq:StateConst_Trans_ValueFunc} subject to \eqref{eq:StateConst_Trans_Constraint}),
    \begin{enumerate}
        \item the stage cost $H^*(s,x,b)$ is convex in $(x,b)$ for each $s\in[t,T]$ if $L^x(s,x)$ in \eqref{eq:Stagecost_Decom} is convex in $x$ for each $s\in[t,T]$ and \eqref{eq:Dynamics_Decom} holds; see Lemma \ref{lemma:StageCost_cvxCondition},
        
        \item the control constraint ($\bar{B}(s)$ in \eqref{eq:Sum_V_set}) is convex in $(x,b)$ for each $s\in[t,T]$ if \eqref{eq:Dynamics_Decom} holds; see Lemma \ref{lemma:CtrlSet_cvxCondition},
        
        \item the dynamics function $-b$ in \eqref{eq:StateConst_Trans_Constraint} is affine in $(x,b)$ for each $s\in[t,T]$ without any assumptions,
        
        \item the terminal cost $g$ has to be assumed as convex in $x\in\R^n$, and the state constraint $c(s,x)$ has to be assumed convex in $x\in\R^n$ for each $s\in[t,T]$.
    \end{enumerate}
    \label{remark:summary_convexity_GHL}
\end{remark}

\subsection{Comparison of convexity conditions}

\begin{table*}[h]
    \centering
    {\renewcommand{\arraystretch}{1.5}%
    \begin{tabular}{c|c|c|c}
         \multicolumn{2}{c|}{ } & the state-constrained  & \multirow{2}{*}{ the generalized Lax formula }\\
         \multicolumn{2}{c|}{ }& optimal control problem &\\\hline
         \multicolumn{2}{c|}{formulation}  & \eqref{eq:def_vartheta} & \eqref{eq:StateConst_Trans_ValueFunc} subject to \eqref{eq:StateConst_Trans_Constraint}\\\hline\hline
         &\multicolumn{3}{c}{costs}\\\cline{2-4}
         &stage cost &   $L^x(s,\cdot)$ is convex in $x$  & $L^x(s,\cdot)$ is convex in $x$ \\
         &$L=L^x (s,x)+L^a(s,a)$&  $L^a(s,\cdot)$ is convex in $a$ &  See Lemma \ref{lemma:StageCost_cvxCondition} \\\cline{2-4}
         &terminal cost& \multirow{2}{*}{convex in $x$}&\multirow{2}{*}{convex in $x$}\\
         &$g$ &&\\\cline{2-4}
         convexity&\multicolumn{3}{c}{constraints}\\\cline{2-4}
         conditions&control constraint & \multirow{2}{*}{convex}& no condition \\
         &$A$&& See Lemma \ref{lemma:CtrlSet_cvxCondition}\\\cline{2-4}
         &dynamics function &\multirow{2}{*}{$f=M(s)x+N(s)a+C(s)$}&$f=M(s)x+\varphi(s,a)$ \\
         &$f(s,x,a)$ & & See Lemma \ref{lemma:StageCost_cvxCondition} and \ref{lemma:CtrlSet_cvxCondition} \\\cline{2-4}
         &state constraint &  \multirow{2}{*}{convex in $x$}& \multirow{2}{*}{convex in $x$}\\
         &$c(s,\cdot)$  &&
    \end{tabular}}
    \caption{Convexity conditions for the temporally discretized state-constrained optimal control problem (\eqref{eq:Numerical_Problem_Cost} subject to \eqref{eq:Numerical_Problem_Const}) and the temporally discretized generalized Lax formula (\eqref{eq:Numerical_HopfLax_Cost} subject to \eqref{eq:Numerical_HopfLax_Const}).}
    \label{tab:class2}
\end{table*}

Table \ref{tab:class2} shows the convexity conditions for the temporally discretized state-constrained optimal control problem (\eqref{eq:Numerical_Problem_Cost} subject to \eqref{eq:Numerical_Problem_Const}) and the temporally discretized generalized Lax formula (\eqref{eq:Numerical_HopfLax_Cost} subject to \eqref{eq:Numerical_HopfLax_Const}).
In this subsection, we still assume that \eqref{eq:Stagecost_Decom} holds.

As summarized in Table \ref{tab:class2}, the temporally discretized generalized Lax formula (\eqref{eq:Numerical_HopfLax_Cost} subject to \eqref{eq:Numerical_HopfLax_Const}) is convex if Condition \ref{cvxCondition:ConvexCondition_HopfLax} holds. In other words, if Conditions \ref{cvxCondition:ConvexCondition_HopfLax} holds, all conditions in Remark \ref{remark:summary_convexity_GHL} are satisfied.
\begin{cvxCondition} \textnormal{\textbf{(The convexity conditions for the temporally discretized generalized Lax formula)}}
    $L$, $g$, $A$, $f$, and $c$ are the stage cost, terminal cost, control constraint, dynamics function, state constraints for the state-constrained problem \eqref{eq:def_vartheta}.
    \begin{enumerate}
        \item $L^x(s,\cdot)$ is convex in $x\in\R^n$ for each $s\in[t,T]$,
        
        \item $g$  is convex in  $x \in \R^n$,
        
        \item $f(s,x,a)=M(s)x+\varphi(s,a)$ in $\R^n \times A$ for each $s\in[t,T]$,
        
        \item $c(s,\cdot)$ is convex in $x\in \R^n$ for each $s\in[t,T]$.
        
    \end{enumerate}
    \label{cvxCondition:ConvexCondition_HopfLax}
\end{cvxCondition}

In comparing Condition \ref{cvxCondition:ConvexCondition_original} and \ref{cvxCondition:ConvexCondition_HopfLax}, Condition \ref{cvxCondition:ConvexCondition_original} always satisfies Condition \ref{cvxCondition:ConvexCondition_HopfLax}. 
In other words, there is a class of problems in which the temporally discretized generalized Lax formula (\eqref{eq:Numerical_HopfLax_Cost} subject to \eqref{eq:Numerical_HopfLax_Const}) is convex even though the temporally discretized state-constrained problem (\eqref{eq:Numerical_Problem_Cost} subject to \eqref{eq:Numerical_Problem_Const}) is non-convex.
In this convexity analysis, we state benefits of the generalized Lax formula in Remark \ref{remark:benefit_GHL}.

\begin{remark} [Benefits of the generalized Lax formula]
~
    Suppose \eqref{eq:Dynamics_Decom} holds.
    For convex temporally discretized generalized Lax formula (\eqref{eq:Numerical_HopfLax_Cost} subject to \eqref{eq:Numerical_HopfLax_Const}), the state-dependent conditions in Condition \ref{cvxCondition:ConvexCondition_original} has to be satisfied, but the control-dependent conditions in Condition \ref{cvxCondition:ConvexCondition_original} are not required.
    In summary, for convex temporally discretized generalized Lax formula,
    \begin{enumerate}
        \item the control-dependent stage $L^a(s,a)$ in \eqref{eq:Stagecost_Decom} is not required to be convex in $a\in A\subset\R^m$ for each $s\in[t,T]$,
        
        \item the control-dependent dynamics $\varphi(s,a)$ in \eqref{eq:Dynamics_Decom} is not required to be affine in $a\in A$ for each $s\in[t,T]$,
        
        \item the control constraint $A$ in \eqref{eq:def_vartheta_ep_const} is not required to be convex.
    \end{enumerate}
    \label{remark:benefit_GHL}
\end{remark}

\section{Examples and Demonstrations}
\label{sec:Examples}

We introduce three examples for the generalized Lax formula and illustrate the benefits of the formula in comparison to the temporally discretized optimal control problem (\eqref{eq:Numerical_Problem_Cost} subject to \eqref{eq:Numerical_Problem_Const}).
For numerical computation, a computer with a 2.8 GHz Quad-Core i7 CPU and 16 GB RAM was used.


\subsection{2D nonlinear vehicle}
\label{sec:example_2D}
We introduce a 2D nonlinear vehicle example where the generalized Lax formula provides a convex problem 
whereas the given optimal control problem is non-convex.
In addition, we know the analytic solution for this problem, which will be compared to the result of the generalized Lax formula and Algorithm \ref{alg:Opt_NewFormulation1}.

A vehicle in the 2D-plane follows the dynamics: $f(s,x,a) = [\cos(a);\sin(a)]$, $A=[-\pi,\pi]$. This vehicle model has speed 1, and the control decides the angle of attack.
The goal of the vehicle is to reach the particular target after 1 sec:
\begin{align}
    &\quad\quad\quad\quad\quad\quad \vartheta(0,x) = \inf_{\alpha} \lVert \mathrm{x}(1) \rVert_2,\label{eq:ex1_prob_def_cost}\\
    &\text{subject to } 
    \begin{cases}
        \dot{\mathrm{x}}(s) = [\cos(\alpha(s));\sin(\alpha(s))], &s\in [0,1], \\
        \mathrm{x}(0) = x,\\
        \alpha(s)\in [-\pi,\pi],  &s\in [0,1],
    \end{cases} 
    \label{eq:ex1_prob_def_const}
\end{align}
where $x$ is the initial state. 
For this example, by Remark \ref{remark:benefit_GHL}, the temporally discretized optimal control problem (\eqref{eq:Numerical_Problem_Cost} subject to \eqref{eq:Numerical_Problem_Const}) is non-convex, but the temporally discretized generalized Lax formula (\eqref{eq:Numerical_HopfLax_Const} subject to \eqref{eq:Numerical_HopfLax_Cost}) is convex.

By the definition of the Hamiltonian and its Fenchel-Legendre transformation in \eqref{eq:Hamiltonian} and \eqref{eq:DEF_Hamil_ConvConj}, for $(s,x,p)\in[0,1]\times\R^2\times\R^2$, $H(s,x,p) = \lVert p \rVert_2$, and for $(s,x,b)\in[0,1]\times\R^2\times\R^2$,
\begin{align*}
    H^*(s,x,b) = \begin{cases}0,& b\in \text{Conv}(B(s,x))=\{b~|~\|b\|_2 \leq 1\}, \\ \infty, & \text{otherwise.}\end{cases}
\end{align*}
Note that $B(s,x)=\{b~|~ \lVert b \rVert_2 =1 \}$ is non-convex in $b$.

The generalized Lax formula provides the following optimal control problem:
\begin{align}
    &\quad\quad\quad\quad\quad\quad\inf_{\beta} \lVert \mathrm{x}(1) \rVert_2, \label{eq:ex1_GLax_cost}\\
    &\text{subject to } 
    \begin{cases}
        \dot{\mathrm{x}}(s) = -\beta(s), &s\in [0,1], \\
        \mathrm{x}(0) = x,\\
        \lVert \beta(s) \rVert_2 \leq 1,  &s\in [0,1].
    \end{cases} 
    \label{eq:ex1_GLax_const}
\end{align}
For this, the temporally discretized generalized Lax formula (\eqref{eq:Numerical_HopfLax_Cost} subject to \eqref{eq:Numerical_HopfLax_Const}) is convex.
We numerically solve this problem using the interior-point method \cite{boyd2004convex} in Matlab and obtain the optimal $\mathrm{x}_*[\cdot]$ and $\beta_*[\cdot]$ sequences under the temporal discretization: $\{t_0=0,...,t_K=1\}$.

Given the numerical optimal control ($\beta_*$) and state ($\mathrm{x}_*$) sequences for the generalized Lax formula, Algorithm \ref{alg:Opt_NewFormulation1} provides the corresponding optimal control signal $\alpha_*^\epsilon$ in \eqref{eq:approx_ctrl} for the given optimal control problem \eqref{eq:ex1_prob_def_cost} subject to \eqref{eq:ex1_prob_def_const}.
To get a numerical $\alpha_*^\epsilon$ in \eqref{eq:approx_ctrl}, we first find $\beta_{i*}[k]\in B(t_k,\mathrm{x}_*[k])$ and $\gamma_i[k]\in[0,1]$ such that
\begin{align}
    \beta_*[k] =  \gamma_1[k] \beta_{1*} [k] + \gamma_2 [k] \beta_{2*}[k],
\end{align}
where  
\begin{align}
\begin{tabular}{cc}
    $\beta_{1*} [k] = \frac{\beta_*[k]}{\lVert \beta_*[k] \rVert_2},$& $\beta_{2*}[k]=-\beta_{1*}[k],$ \\
    $\gamma_1[k]=\frac{1+ \lVert \beta_*[k] \rVert_2 }{2},$ & $\gamma_2 [k] =\frac{1- \lVert \beta_*[k] \rVert_2 }{2}.$
\end{tabular}
\end{align}
for $\beta_*[k]\in\text{Conv}(B(t_k,\mathrm{x}_*[k]))$ as in Lemma \ref{lemma:lemma6}.
Note that \eqref{eq:LinearComb_cost} also holds for $\beta_{i*}[k]$ and $\gamma_{i}[k]$ by Corollary \ref{corollary:NewStagecost_NoControlInput} since $H^*(s,x,b)= L(s,x,a)= 0$ for all $(s,x,a,b)\in[0,T]\times\R^n\times\R^m\times\R^n$.

Thus, the approximate-optimal control signal $\alpha^\epsilon_*$ in \eqref{eq:approx_ctrl} is
\begin{align*}
    \alpha^\epsilon_*(s) =\begin{cases} \arctan \frac{\beta_{1*} [k,2]}{\beta_{1*} [k,1]}, & s\in[t_k, t_k+\Delta t_k \gamma_1[k]), \\ \arctan \frac{\beta_{2*} [k,2] }{\beta_{2*} [k,1] }, & s\in[t_k+\Delta t_k \gamma_1[k],t_{k+1}), \end{cases}
\end{align*}
where $\Delta t_k = t_{k+1}-t_k$ and $\beta_{i*}[k]=[\beta_{i*}[k,1];\beta_{i*}[k,2]]$, $i=1,2$.

\begin{figure*}[h!]
\centering
\begin{tabular}{ccc}
    \includegraphics[trim = 0mm 0mm 0mm 0mm, clip, height=0.17\textwidth]{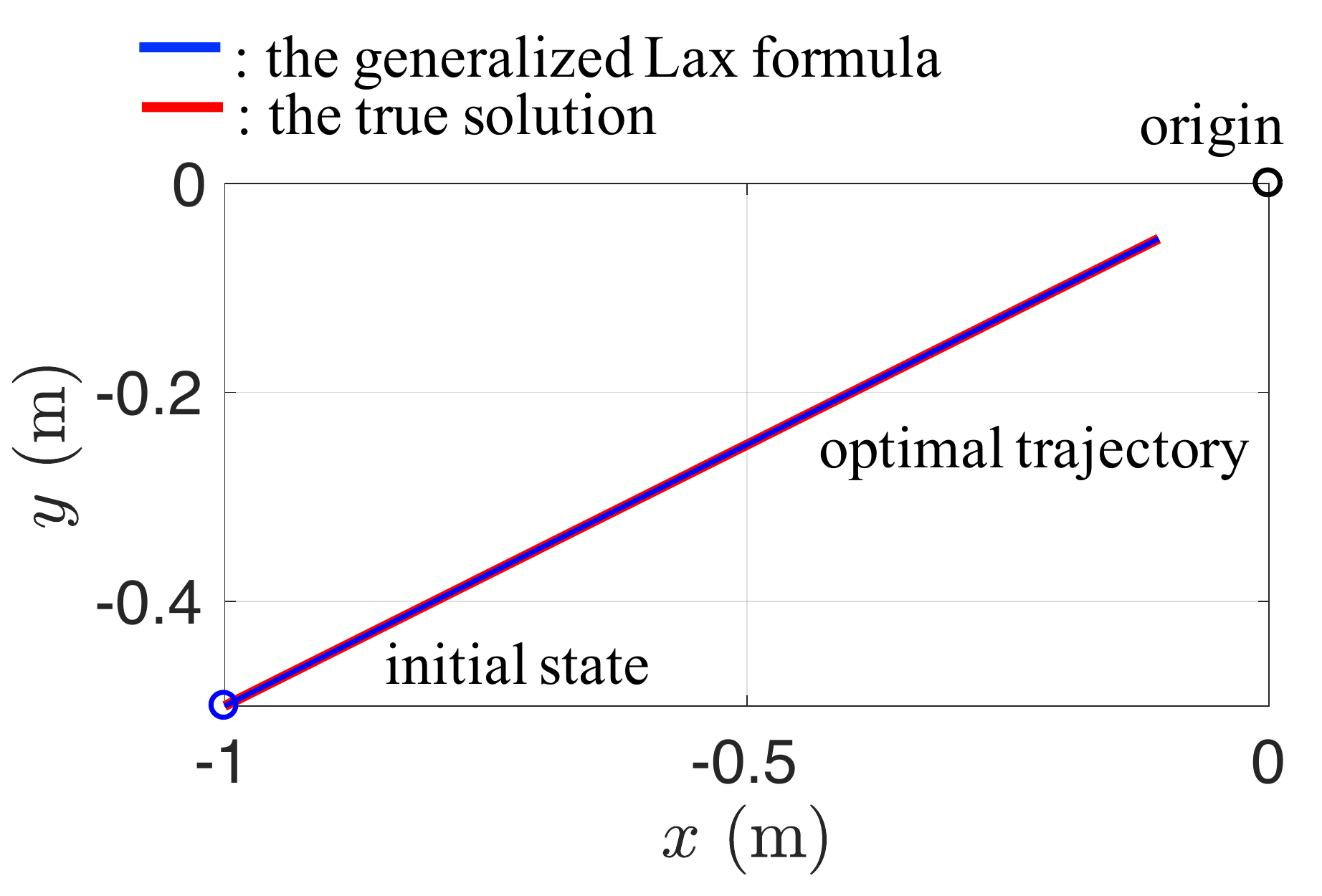} & \includegraphics[trim = 0mm 0mm 0mm 0mm, clip, height=0.16\textwidth]{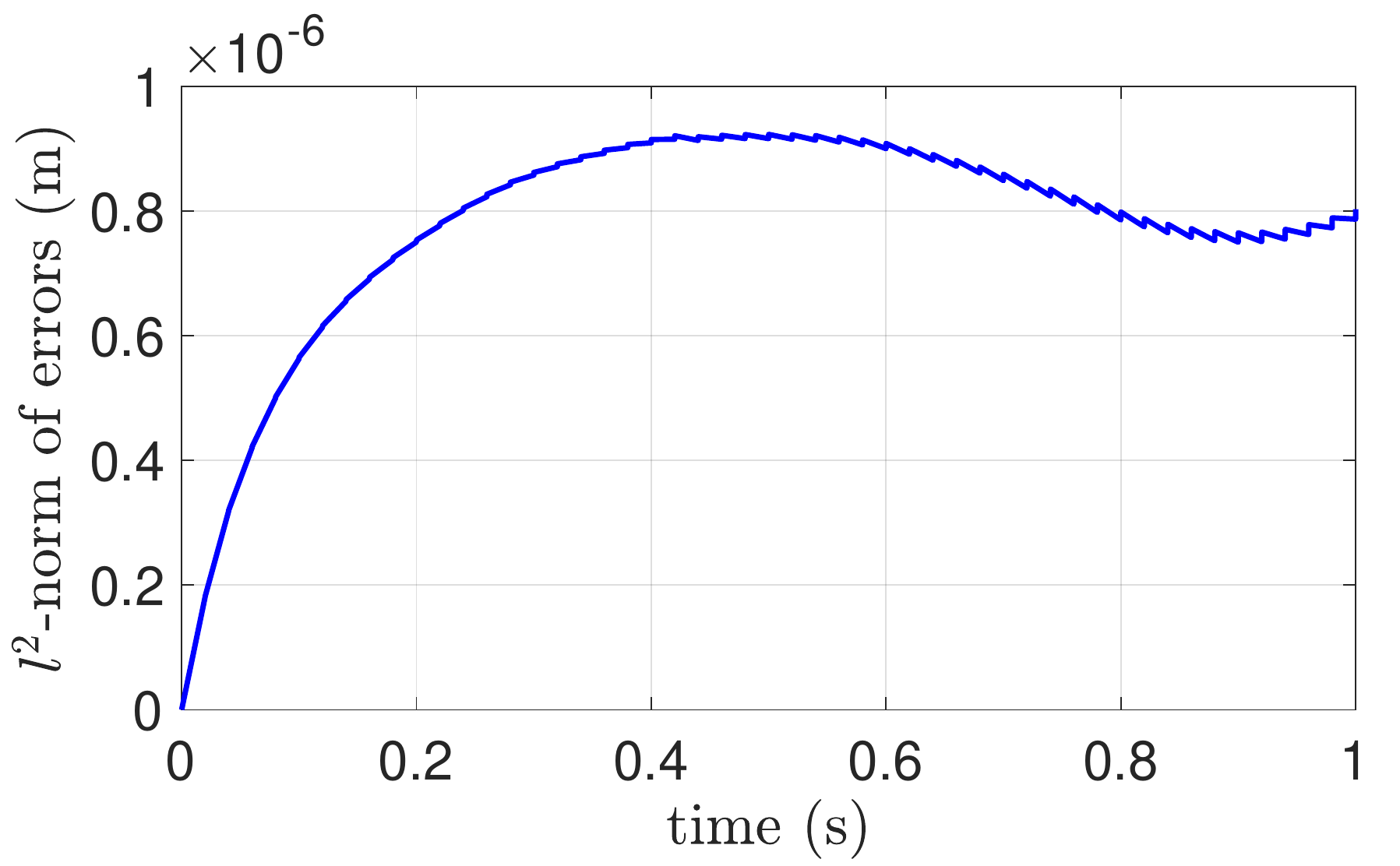} & \includegraphics[trim = 0mm 0mm 0mm 0mm, clip, height=0.17\textwidth]{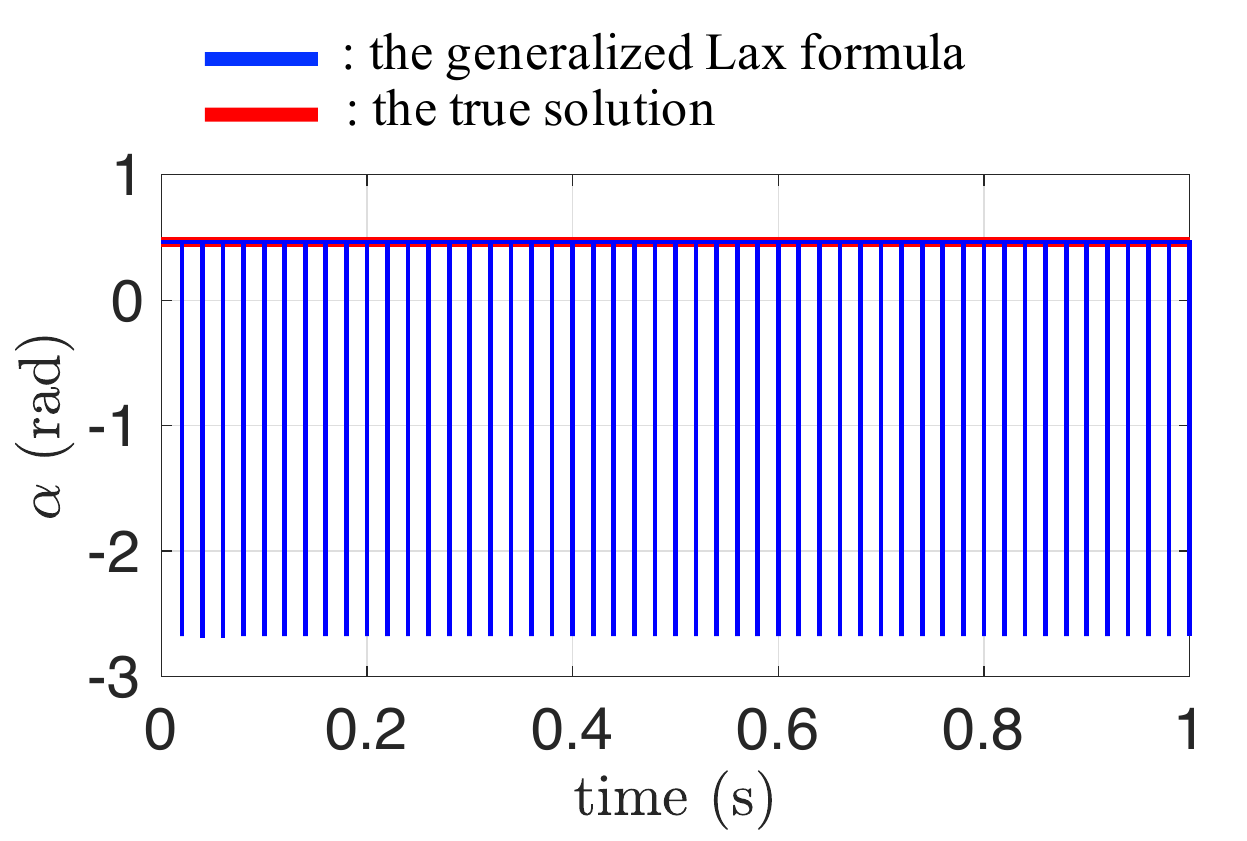}  \\
    (a) & (b) & (c)
\end{tabular}
\caption{ A state-unconstrained optimal control problem for the two-dimensional vehicle in Section \ref{sec:example_2D}. (a) Comparison of the optimal state trajectory using the generalized Lax formula (blue) and the analytic solution (red). Since the two trajectories are very close and hard to discriminate, we show, in (b), the $l^2$-norm of the errors between the analytic solution and the numerical solution by the generalized Lax formula. (c) Comparison between the analytic optimal control signal ($u_*(\cdot)\equiv \arctan(0.5)$) in red and the optimal control signal using the generalized Lax formula with Algorithm  \ref{alg:Opt_NewFormulation1} in blue. Most of the time, the optimal control signal derived by the generalized Lax formula are very close to the true solution and shows a frequent control switching, induced by numerical error. This control switching issue is discussed in the later part of Section \ref{sec:NewFormulation}.  }
\label{fig:Example_car2d}
\end{figure*}

The computation time for Algorithm \ref{alg:Opt_NewFormulation1} is 1.29 s.

Figure \ref{fig:Example_car2d} illustrates the numerical optimal state trajectory and control signal derived by the generalized Lax formula and Algorithm \ref{alg:Opt_NewFormulation1} in comparison to the analytic solution for given initial state $x=[-1;-0.5]$.
As shown in Figure \ref{fig:Example_car2d} (b), the $l^2$-norm error of the state trajectory over time is bounded by $1\times10^{-6}$.
The optimal control signal is analytically $\alpha_*\equiv \arctan(0.5)$ for all time.
As shown in Figure \ref{fig:Example_car2d} (c), the numerical optimal control signal $\alpha_*^\epsilon$ is $\arctan(0.5)$ for most of the time in $[0,1]$.

\subsection{Formation control of multiple nonlinear vehicles}
\label{sec:example_12D}
We introduce a 12D nonlinear example, a formation control for multiple agents whose dynamics are nonlinear:
\begin{align}
    f(s,\mathrm{x}^l,\alpha^l(s))=\left[
    \begin{array}{l}
        \dot{\mathrm{x}}^{l} (s,1)  \\
        \dot{\mathrm{x}}^{l} (s,2) \\
        \dot{\mathrm{x}}^{l} (s,3) \\
        \dot{\mathrm{x}}^{l} (s,4) 
    \end{array}
    \right]=
    \left[
    \begin{array}{l}
        \mathrm{x}^{l} (s,2)\\
        \alpha^l (s,1) \cos (\alpha^l (s,2)) \\
        \mathrm{x}^{l} (s,4) \\
        \alpha^{l} (s,1) \sin (
        \alpha^{l} (s,2))
    \end{array}
    \right],
    \label{eq:multiagentDyn}
\end{align}
where $l\in\{1,2,3\}$ is the agent index, $\mathrm{x}^{l}(s,1)$ and $\mathrm{x}^{l}(s,2)$ are horizontal position and velocity in the 2D space, $\mathrm{x}^{l}(s,3)$ and $\mathrm{x}^{l}(s,4)$ are vertical position and velocity in the 2D space, and $\alpha^{l}(s,1)$ and $\alpha^{l}(s,2)$ are the magnitude of the acceleration and the angle of agent $l$, respectively, at time $s\in[t,T]$. 
For three agents, the dimension of the state is twelve, and the dimension of the control is six.

We define an optimal control problem where three agents approach the goal point with the right-triangular-shaped formation:
\begin{align}
    &\inf_{\alpha} \int_0^{10}  \max \bigg\{  \bigg\Vert \begin{bmatrix}\mathrm{x}^1(s,1)\\\mathrm{x}^1(s,3)\end{bmatrix}-\begin{bmatrix}\mathrm{x}^{1,r} (s,1)\\\mathrm{x}^{1,r} (s,3)\end{bmatrix}\bigg\Vert_2, \notag\\
    & \quad\quad \bigg\Vert \begin{bmatrix}\mathrm{x}^2(s,1)\\\mathrm{x}^2(s,3)\end{bmatrix}-\begin{bmatrix}\mathrm{x}^1(s,1)\\\mathrm{x}^1(s,3)\end{bmatrix}- d^r \bigg\Vert_2 , \notag\\
    & \quad\quad \bigg\Vert \begin{bmatrix}\mathrm{x}^3(s,1)\\\mathrm{x}^3(s,3)\end{bmatrix} - h\bigg(\begin{bmatrix}\mathrm{x}^1(s,1)\\\mathrm{x}^1(s,3)\end{bmatrix} ,\begin{bmatrix}\mathrm{x}^2(t,1)\\\mathrm{x}^2(t,3)\end{bmatrix}\bigg)\bigg\Vert_2 \bigg\} ds
    \label{eq:exFormationCostPrimal}\\
     &\text{subject to } \begin{cases}
        \eqref{eq:multiagentDyn},  \\
        \mathrm{x}^l (0) = x^l , \\
        \alpha^l(s,1)\in [-1,3],~~
        \alpha^l(s,2)\in [-\frac{\pi}{6},\frac{\pi}{6}], 
    \end{cases} 
    \label{eq:exFormationConstPrimal}
\end{align}
for $s\in [0,T]$, $l=1,2,3$,
where $\mathrm{x}^{1,r}(s,1)=2s$, $\mathrm{x}^{1,r}(s,3)=0$, $d^r=[-\sqrt{3};1]\in\R^2$, and, for $w_1,w_2\in\R^2$,
\begin{align*}
    h\left(w_1,w_2\right)\coloneqq\begin{bmatrix}\frac{1}{2} & \frac{\sqrt{3}}{2} \\ -\frac{\sqrt{3}}{2} &\frac{1}{2}\end{bmatrix}w_1 +\begin{bmatrix}\frac{1}{2} & -\frac{\sqrt{3}}{2} \\ \frac{\sqrt{3}}{2} &\frac{1}{2}\end{bmatrix} w_2.
\end{align*}
Agent 1 is the leader that tracks the reference trajectory $\mathrm{x}^{1,r}$, for which the first term of the state cost is designed. 
Agent 2 is following Agent 1 with $d^r$-offset, designed in the second term of the stage cost.
Agent 3 is making the right-triangular formation, for which the third term of the stage cost is designed.

By Remark \ref{remark:benefit_GHL}, the temporally  discretized optimal control problem (\eqref{eq:Numerical_Problem_Cost} subject to \eqref{eq:Numerical_Problem_Const}) is non-convex, but the temporally discretized generalized Lax formula (\eqref{eq:Numerical_HopfLax_Cost} subject to \eqref{eq:Numerical_HopfLax_Const}) is convex.

\begin{figure}[t!]
\centering
\begin{tabular}{cc}
     \includegraphics[trim = 0mm 0mm 0mm 0mm, clip, width=0.23\textwidth]{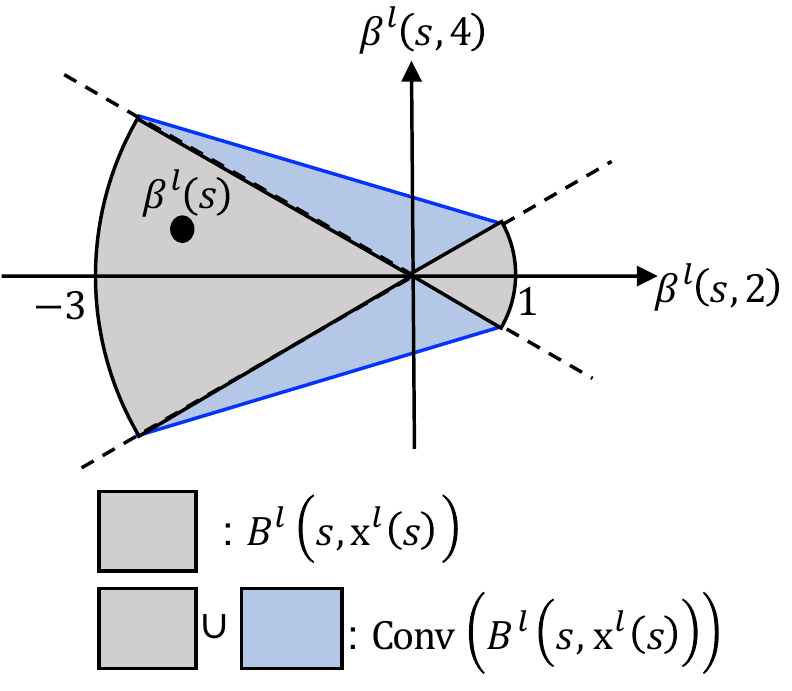} & \includegraphics[trim = 0mm 0mm 0mm 0mm, clip, width=0.23\textwidth]{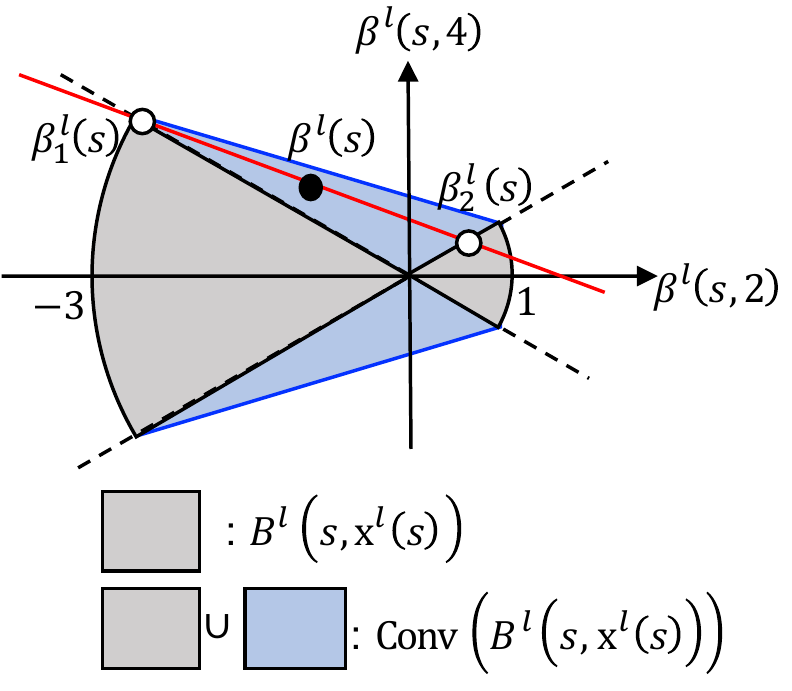}  \\
    (a) & (b) 
\end{tabular}
\caption{ The control constraint for the 12D formation control
problem in Section \ref{sec:example_12D}. $l$ is the index of the agents in \{1,2,3\}.
The equations of the two dotted lines for the both figures are $\sqrt{3}\beta^l(s,4) \pm  \beta^l(s,2)=0$.
These figures illustrate how to find a finite $\beta^l_{i}(s)$ satisfying the control decomposition in Lemma \ref{lemma:lemma6} for $(s,\mathrm{x}^l(s))$ by dividing into two cases.
(a) If $\beta^l(s)\in B^l(s,\mathrm{x}^l(s))$, we do not need the control decomposition in Lemma \ref{lemma:lemma6}. 
(b) If $\beta^l(s)\notin B^l(s,\mathrm{x}^l(s))$, there are multiple choices of $\beta_{i*}^l(s)$. Among the choices, we select $\beta^l_{1}(s)=[-\mathrm{x}^l(s,2);-\frac{3\sqrt{3}}{2};-\mathrm{x}^l(s,4);\pm \frac{3}{2}]$ and find $\beta_{2}^l (s)$: the intersection of the red line connecting $\beta^l (s)$ and $\beta_{1}^l(s)$, and one of the two dotted lines.
}
\label{fig:Example_formationCtrl_CtrlConst}
\end{figure}

By Corollary \ref{corollary:NewStagecost_NoControlInput}, the stage cost for the generalized Lax formula ($H^*$) is equal to the stage cost of the given problem ($L$) since the control-dependent stage cost ($L^a$) in \eqref{eq:Stagecost_Decom} is zero.
To derive the control constraint ($\text{Conv}(B(s,\mathrm{x}(s)))$) for the generalized Lax formula, we use Lemma \ref{lemma:Lemma1}.  
Denote $\mathrm{x}(s)=[\mathrm{x}^1(s);\mathrm{x}^2(s);\mathrm{x}^3(s)]$, $\beta(s)=[\beta^1(s);\beta^2(s);\beta^3(s)]$ and $\beta^l(s)=[\beta^l(s,1);\beta^l(s,2);\beta^l(s,3);\beta^l(s,4)]$, $l=1,2,3$.
By the definition of $B(s,\mathrm{x}(s))$ in \eqref{eq:Trans_CtrlSet}, 
\begin{align}
    \begin{split}
     &B(s,\mathrm{x}(s)) = B^1(s,\mathrm{x}^1(s)) \times B^2(t,\mathrm{x}^2(s)) \\
     &\quad\quad\quad\quad\quad\quad\quad\quad\quad\quad\quad\quad\quad\times B^3(s,\mathrm{x}^3(s))\in\mathbb{R}^{12},
    \end{split}
    \\
    \begin{split}
    & B^l(s,\mathrm{x}^l(s)) = \{ \beta^l(s,1) = -\beta^l (s,2) ,\beta^l(s,3)=-\beta^l(s,4)   \} \cap \\
    &\quad\quad\quad \bigg[ \bigg\{ \bigg\lVert \begin{bmatrix}\beta^l(s,2)\\\beta^l(s,4)\end{bmatrix} \bigg\rVert_2 \leq 3, \begin{bmatrix}1&\sqrt{3}\\1&-\sqrt{3}\end{bmatrix}\begin{bmatrix}\beta^l(s,2)\\\beta^l(s,4)\end{bmatrix}\leq 0 \bigg\} \\
    &\quad\quad \cup \bigg\{ \bigg\lVert \begin{bmatrix}\beta^l(s,2)\\\beta^l(s,4)\end{bmatrix} \bigg\rVert_2 \leq 1, \begin{bmatrix}-1&-\sqrt{3}\\-1&\sqrt{3}\end{bmatrix}\begin{bmatrix}\beta^l(s,2)\\\beta^l(s,4)\end{bmatrix}\leq 0\bigg\}\bigg]
    \end{split}
\end{align}
\hspace*{-0.17cm}shown in the grey in Figure \ref{fig:Example_formationCtrl_CtrlConst} (a).
Then, $\text{Conv}(B(s,x))$ is derived in the last five lines in \eqref{eq:Example_formationCtrl_GHL_Const} and also illustrated in Figure \ref{fig:Example_formationCtrl_CtrlConst}. 
Note that $B(s,\mathrm{x}(s))$ is non-convex, but $\text{Conv}(B(s,\mathrm{x}(s)))$ is convex in $b$.

The generalized Lax formula in Theorem \ref{thm:GHopfLax} provides the following optimal control problem:
\begin{align}
\begin{split}
    &\inf_{\alpha} \int_0^{10}  \max \bigg\{  \bigg\Vert \begin{bmatrix}\mathrm{x}^1(s,1)\\\mathrm{x}^1(s,3)\end{bmatrix}-\begin{bmatrix}\mathrm{x}^{1,r} (s,1)\\\mathrm{x}^{1,r} (s,3)\end{bmatrix}\bigg\Vert_2, \\
    & \quad\quad \bigg\Vert \begin{bmatrix}\mathrm{x}^2(s,1)\\\mathrm{x}^2(s,3)\end{bmatrix}-\begin{bmatrix}\mathrm{x}^1(s,1)\\\mathrm{x}^1(s,3)\end{bmatrix}- d^r \bigg\Vert_2 , \\
    & \quad\quad \bigg\Vert \begin{bmatrix}\mathrm{x}^3(s,1)\\\mathrm{x}^3(s,3)\end{bmatrix} - h\bigg(\begin{bmatrix}\mathrm{x}^1(s,1)\\\mathrm{x}^1(s,3)\end{bmatrix} ,\begin{bmatrix}\mathrm{x}^2(t,1)\\\mathrm{x}^2(t,3)\end{bmatrix}\bigg)\bigg\Vert_2 \bigg\} ds
    \end{split}\label{eq:Example_formationCtrl_GHL_Cost}\\
    &\text{subject to } \begin{cases}
        \dot{\mathrm{x}}^l (s)=-\beta^l(s),~~
        \mathrm{x}^l (0) = x^l,\\
        \beta^l(s,1) = - \mathrm{x}^l(s,2),~~
        \beta^l(s,3) = - \mathrm{x}^l(s,4),\\
        -\beta^l(s,2)-\sqrt{9-(\beta^l(s,4))^2}\leq 0,\\
        \beta^l(s,2)-\sqrt{1-(\beta^l(s,4))^2 } \leq 0,\\
        \frac{1}{2\sqrt{3}}\beta^l(s,2)+\beta^l(s,4)-\frac{3}{4}\leq 0,\\
        \frac{1}{2\sqrt{3}}\beta^l(s,2)-\beta^l(s,4)-\frac{3}{4}\leq 0,
    \end{cases} 
    \label{eq:Example_formationCtrl_GHL_Const}
\end{align}
for $s\in [0,10]$, $l=1,2,3$.
Following the temporal discretization on $\{t_0=0,...,t_K=10\}$ as in \eqref{eq:Numerical_HopfLax_Cost} subject to \eqref{eq:Numerical_HopfLax_Const}, 
we have a convex problem, and gradient-based methods provide a global optimal solution.
For numerical optimization, the interior-point method \cite{boyd2004convex} is utilized, and we denote the optimal state sequence $\mathrm{x}_*[\cdot]$ and control sequence $\beta_*[\cdot]$ for the temporally discretized generalized Lax formula. 
Also, $\mathrm{x}_*[k]=[\mathrm{x}_*^1[k];\mathrm{x}^2_*[k];\mathrm{x}^3_*[k]]$ and $\beta_*[k]=[\beta_*^1[k];\beta^2_*[k];\beta^3_*[k]]$.

Similar to Section \ref{sec:example_2D}, using Algorithm \ref{alg:Opt_NewFormulation1}, we will get the numerical optimal control signal ($\alpha_*^\epsilon:[0,10]\rightarrow\R^6$) in \eqref{eq:approx_ctrl} and state trajectory ($\mathrm{x}_*^\epsilon:[0,10]\rightarrow\R^{12}$) in \eqref{eq:approx_traj}.
To numerically get $\alpha_*^\epsilon$ in \eqref{eq:approx_ctrl}, we need to find $\beta_{i*}^l[k]\in B^l(t_k,\mathrm{x}_*^l [k])$ and $\gamma_i^l[k]\in[0,1]$ ($l=1,2,3$) such that
\begin{align}
    \beta_*^l[k] =  \gamma_{1}^l[k] \beta_{1*}^l [k] + \gamma_{2}^l[k] \beta_{2*}^l[k]
\end{align}
for $\beta_*^l[k]\in\text{Conv}(B^l (t_k,\mathrm{x}_*^l[k]))$ as in Lemma \ref{lemma:lemma6}.
By substituting $t_k$ for $s$ and $\mathrm{x}^l_*[k]$ for $\mathrm{x}^l(s)$, Figure \ref{fig:Example_formationCtrl_CtrlConst} graphically illustrates two cases to find $\beta_{i*}^l[k]$ and $\gamma_{i}^l[k]$: if $\beta_{*}^l[k]$ is in $B^l(t_k,\mathrm{x}_*^l[k])$ or not. 
For each case, the mathematical expression for $\beta_{i*}^{l }$ can be found in Figure \ref{fig:Example_formationCtrl_CtrlConst}.
Note that \eqref{eq:LinearComb_cost} also holds for $\beta_{i*}^l[k]$ and $\gamma_{i}^l[k]$ since $H^*(s,x,b)= L(s,x,a)$ for all $(s,x,a,b)\in[0,T]\times\R^n\times\R^m\times\R^n$ by Corollary \ref{corollary:NewStagecost_NoControlInput}.
Then, we can design
\begin{align}
    \alpha^{\epsilon l}_*(s) = \begin{cases}\alpha^l_{1*}[k],&s\in[t_k,t_k+\Delta t_k \gamma_1^l[k]),\\\alpha^l_{2*}[k],&s\in[t_k+\Delta t_k \gamma_1^l[k],t_{k+1}),\end{cases}
\end{align}
where $\alpha^l_{1*}[k]$ and $\alpha^l_{2*}[k]$ satisfy
\begin{align}
    \begin{split}
        &\beta^l_{1*}[k]=-f(t_k,\mathrm{x}^l_{*}[k],\alpha^l_{1*}[k]),\\
        &\beta^l_{2*}[k]=-f(t_k,\mathrm{x}^l_{*}[k],\alpha^l_{2*}[k]),
    \end{split}
\end{align}
and finally compute $\mathrm{x}_*^{\epsilon l}$ solving \eqref{eq:multiagentDyn} for $\alpha^{\epsilon l}_*$ for each agent.

\begin{figure}[t!]
\centering
\begin{tabular}{c}
\includegraphics[trim = 0mm 0mm 0mm 0mm, clip, width=0.48\textwidth]{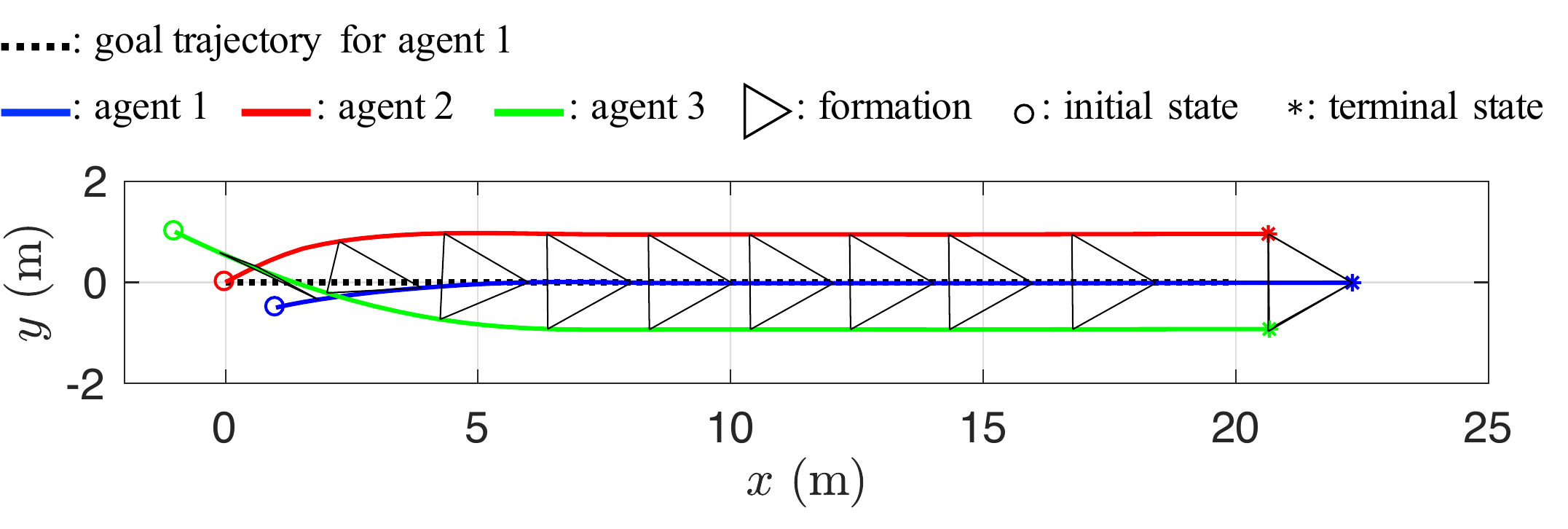}
\\(a)\\
\includegraphics[trim = 0mm 0mm 0mm 0mm, clip, width=0.48\textwidth]{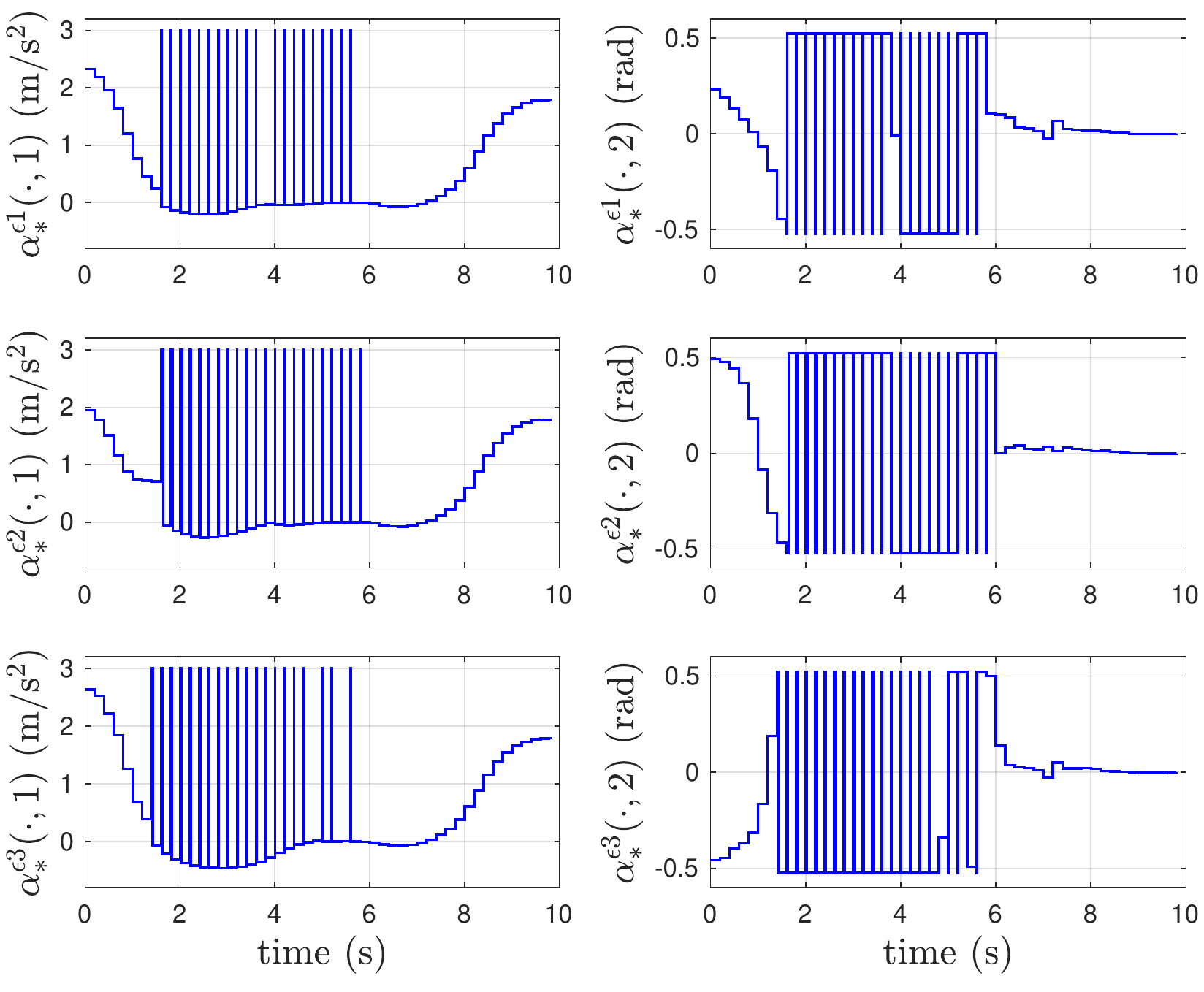}
\\(b)
\end{tabular}
\caption{ The numerical results for the 12D formation control in Section \ref{sec:example_12D} using the generalized Lax formula and Algorithm \ref{alg:Opt_NewFormulation1}. (a) The optimal state trajectories for all agents in the 2D-space are shown with triangular formation at each second. (b) The optimal control signals for all agents are shown. We observe that the control switching is caused by the additional discretization step in Algorithm \ref{alg:Opt_NewFormulation1}. To overcome this issue, we provide practical suggestions without proof.
}
\label{fig:Example_formationCtrl}
\end{figure}

Figure \ref{fig:Example_formationCtrl} illustrates the numerical optimal state trajectory ($\mathrm{x}_*^\epsilon=[\mathrm{x}_*^{\epsilon 1},\mathrm{x}_*^{\epsilon 2},\mathrm{x}_*^{\epsilon 3}]$) and control signal ($\alpha_*^\epsilon=[\alpha_*^{\epsilon 1},\alpha_*^{\epsilon 2},\alpha_*^{\epsilon 3}]$) computed by the generalized Lax formula and Algorithm \ref{alg:Opt_NewFormulation1}. 
This result is theoretical guaranteed to be a globally optimal solution even though the given optimal control problem (\eqref{eq:exFormationCostPrimal} subject to \eqref{eq:exFormationConstPrimal}) is non-convex.
The computation time for Algorithm \ref{alg:Opt_NewFormulation1} is 152 s. 
On the other hand, it is not realistic to numerically solve the HJB PDE for the given problem using the level-set method \cite{Osher02} and the fast marching method \cite{sethian1996fast} due to their exponential complexity in computation when the dimension of the state is more than five.

\subsection{Gear system (hybrid system)}
\label{sec:example_gear}

We introduce a 4D gear system as an example of hybrid systems where the coordinates of the control are constrained in both continuous and discrete space.
Since the set defined in discrete space is always non-convex, the temporally discretized state-constrained optimal control problem (\eqref{eq:Numerical_Problem_Cost} subject to \eqref{eq:Numerical_Problem_Const}) is non-convex regardless of the convexity of the cost in \eqref{eq:Numerical_Problem_Cost}.

To handle optimization problems with the discrete constraints, mixed-integer programming (MIP) is typically used. However, MIP is a non-convex programming.
In this example, we will show that the generalized Lax formula converts the discrete control constraint ($A$ in \eqref{eq:def_vartheta_ep_const}) to the continuous control constraint ($\text{Conv}(B(s,x))$ in \eqref{eq:StateConst_Trans_Constraint}) without any approximation.

\begin{figure}[t!]
\centering
\begin{tabular}{cc}
     \includegraphics[trim = 0mm 0mm 0mm 0mm, clip, width=0.16\textwidth]{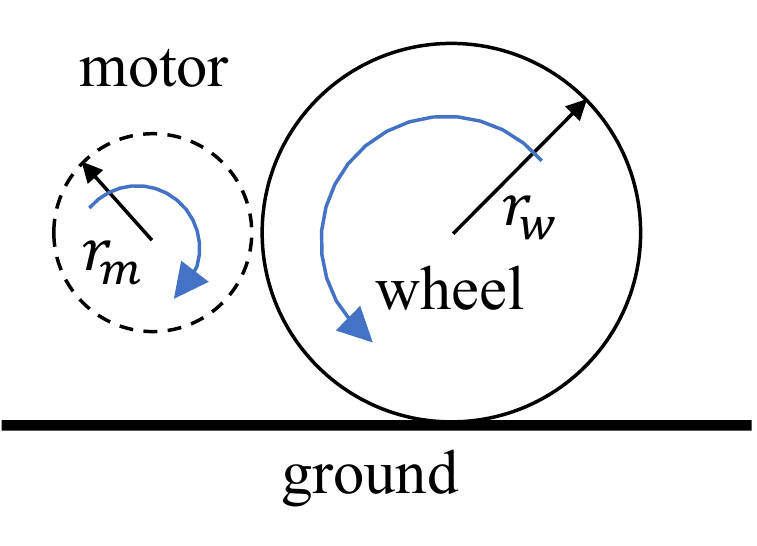} & \includegraphics[trim = 0mm 0mm 0mm 0mm, clip, width=0.30\textwidth]{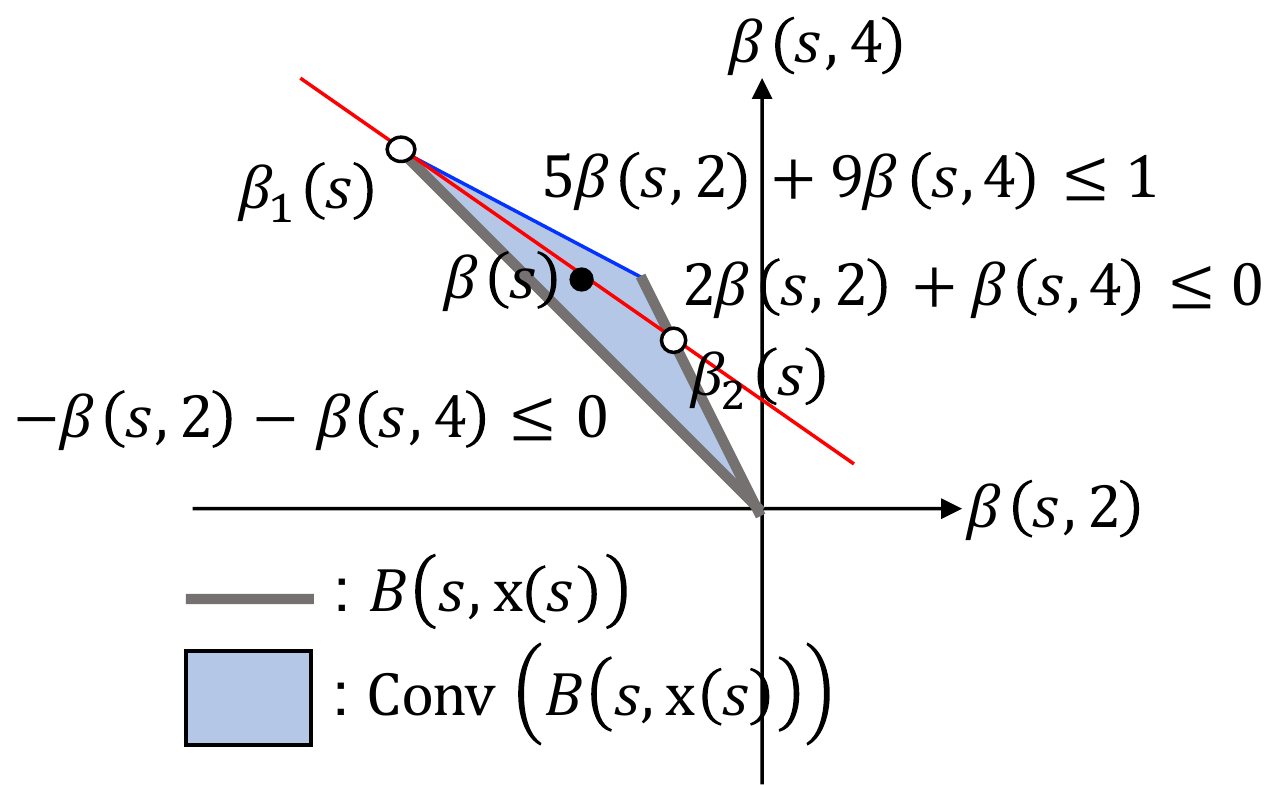}  \\
    (a) & (b) 
\end{tabular}
\caption{(a) This illustration shows the gear system in Section \ref{sec:example_gear} in which $r_m$ varies. (b) The control constraint for the generalized Lax formula is $\beta(s,1)=-\mathrm{x}(s,2)$, $\beta(s,3)=-\mathrm{x}(s,4)$, and the inside of the blue region for $(\beta(s,2),\beta(s,4))$.
By Lemma \ref{lemma:lemma6}, $\beta(s)$ in $\text{Conv}(B(s,\mathrm{x}(s)))$ is linearly decomposed into $\beta_1(s)$ and $\beta_2(s)$. Among multiple ways to choose $\beta_1(s)$ and $\beta_2(s)$, $(-\mathrm{x}(s,2);-\frac{1}{4};-\mathrm{x}(s,4);\frac{1}{4})$ is chosen for $\beta_1(s)$, and the intersection of the red line connecting $\beta(s)$ and $\beta_1(s)$ and the line $2\beta_2(s,2)+\beta_2(s,4)=0$ is chosen for $\beta_2(s)$.
}
\label{fig:Example_gear}
\end{figure}

\begin{figure}[t!]
\centering
\begin{tabular}{c}
     \includegraphics[trim = 0mm 0mm 0mm 0mm, clip, width=0.40\textwidth]{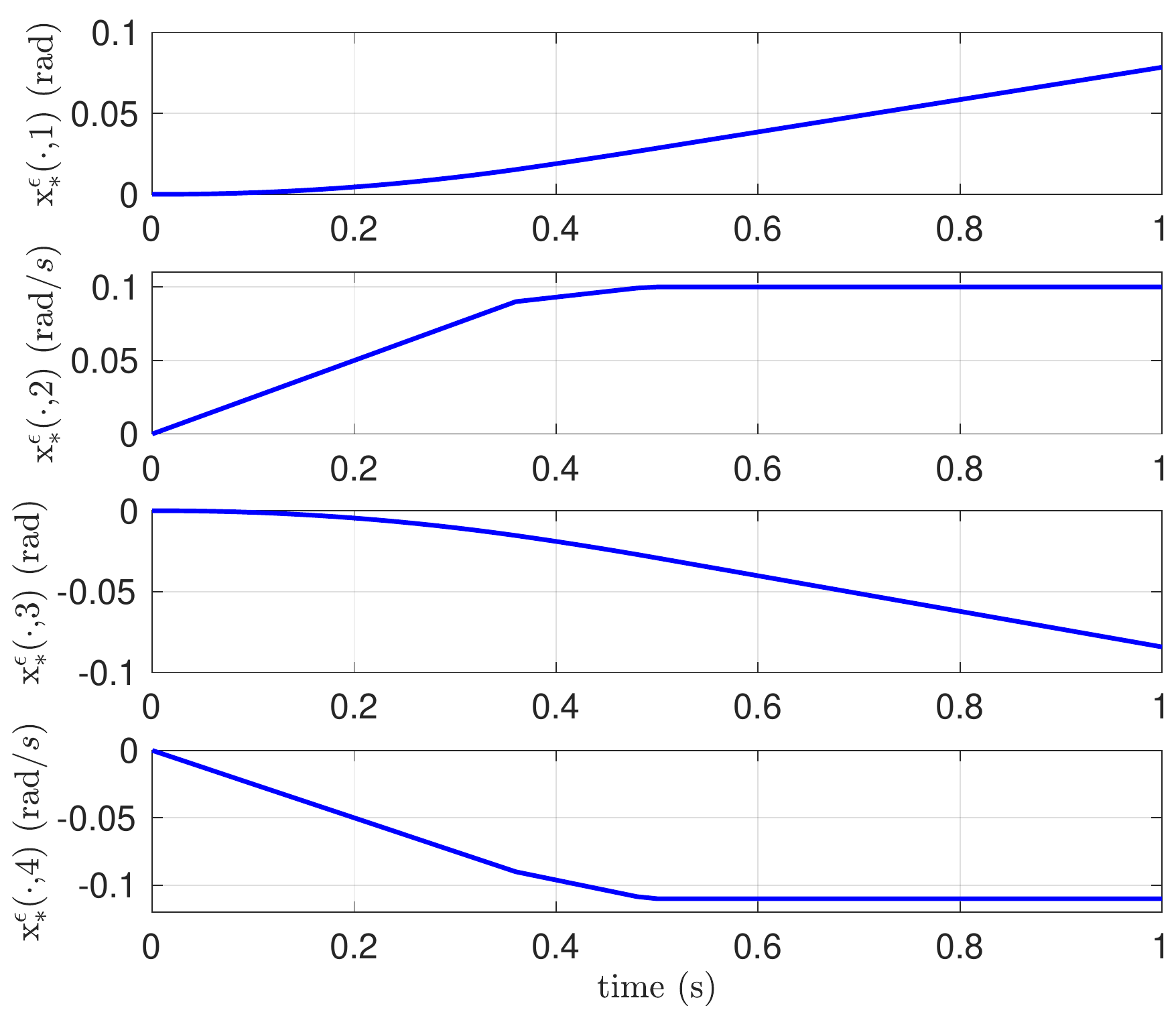} 
     \\(a)\\
     \includegraphics[trim = 0mm 0mm 0mm 0mm, clip, width=0.40\textwidth]{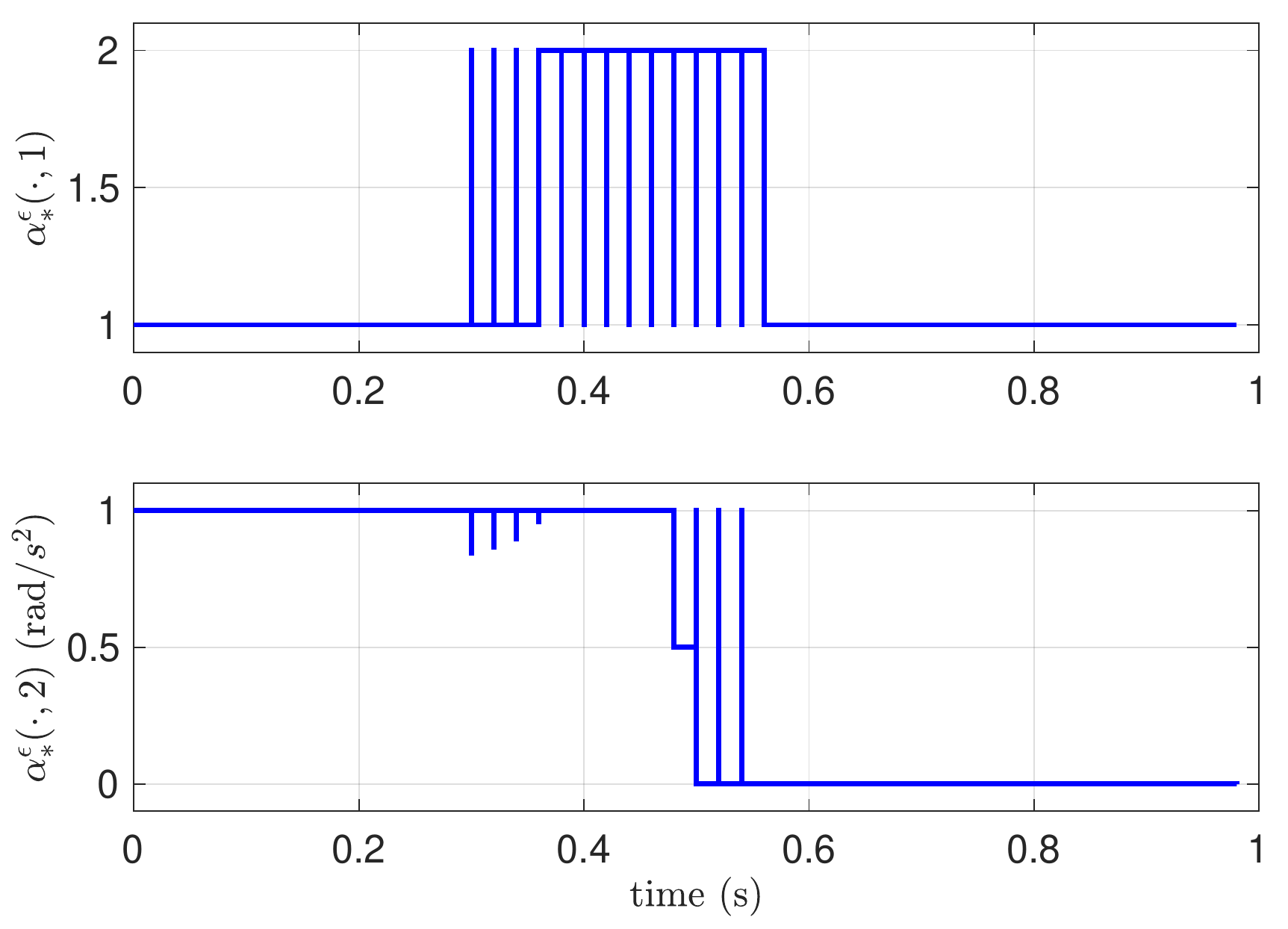}  \\
    (b) 
\end{tabular}
\caption{ For the state-constrained optimal control problem with the gear system, (a) the optimal state trajectories and (b) the optimal control signals are obtained by the generalized Lax formula.
Similar to the previous two examples, we observe the control switching in short time, which is caused by the additional discretization in Algorithm \ref{alg:Opt_NewFormulation1}. 
}
\label{fig:Example_gear_result}
\end{figure}


As described in Figure \ref{fig:Example_gear} (a), the motor and the wheel are connected via a gear system, and the dynamics are following:
\begin{align}
\begin{split}
    & I_m \ddot{\mathrm{q}}_m (s) = \tau_m (s) -r_m(s) f(s),  \\
    & I_w \ddot{\mathrm{q}}_w (s) = -r_w f(s) + r_w f_c (s),\\
    & (m_m + m_w) \ddot{\mathrm{q}}_w (s) = f_c(s),
\end{split}
\end{align}
where $\mathrm{q}_m(s)$ and $\dot{\mathrm{q}}_m(s)$ are the angle and angular velocity of the motor at $s\in[t,T]$, $\mathrm{q}_w(s)$ and $\dot{\mathrm{q}}_w(s)$ are the angle and angular velocity of the wheel at $s\in[t,T]$, $I_m$ and $I_w$ are the inertia of the motor and the wheel, respectively, $m_m$ and $m_w$ are the mass of the motor and the wheel, respectively, $\tau_m$ is the torque by the motor, $r_m$ and $r_w$ are radius of the motor and the wheel, respectively, $f$ is the contact force between the motor and the wheel, and $f_c$ is the friction force from the ground to the wheel. 
We suppose that the radius of the motor changes and determine the gear ratio. 
We assume that the wheel does not slip on the ground. 
By this assumption, $f_c (s) = -(m_m + m_w) r_w \ddot{\mathrm{q}}_w(s)$,
and the dynamics becomes
\begin{align}
    \bigg[I_m + \Big(\frac{r_m(s)}{r_w}\Big)^2 I_w +r_m^2 (s) (m_m+m_w) \bigg]\ddot{\mathrm{q}}_m (s) = \tau_m (s).
\end{align}

We denote $\mathrm{x}(s)=[\mathrm{x}(s,1);\mathrm{x}(s,2); \mathrm{x}(s,3);\mathrm{x}(s,4)]=[\mathrm{q}_m(s);\dot{\mathrm{q}}_m(s); \mathrm{q}_w(s); \dot{\mathrm{q}}_w(s)]$, $\alpha(s)= [\alpha(s,1);\alpha(s,2)]=[\frac{r_m (s)}{r_w}; \tau_m(s)]$ for $s\in[t,T]$. 
Set $I_m=I_w=m_m=m_w=r_w=1$. Then,
\begin{align}
\begin{split}
    &f(s,\mathrm{x}(s),\alpha(s))\\
    &\quad=\left[
    \begin{array}{l}
        \dot{\mathrm{x}} (s,1) \\
        \dot{\mathrm{x}} (s,2) \\
        \dot{\mathrm{x}} (s,3)  \\
        \dot{\mathrm{x}} (s,4) 
    \end{array}
    \right] = \left[
    \begin{array}{l}
        \mathrm{x}(s,2) \\
        \frac{1}{c_1  + c_2 \alpha(s,1)^2 }\alpha(s,2) \\
        \mathrm{x}(s,4) \\
        - \frac{\alpha(s,1)}{c_1 + c_2 \alpha(s,1)^2 }\alpha(s,2)
    \end{array}
    \right],
    \end{split}
    \label{eq:ex_grear_Dyn1}
\end{align}
where $c_1 = I_m=1$, $c_2 =I_w + r_w^2 (m_m + m_w) =3 $.

We solve the state-constrained optimal control problem where the wheel wants to go further to the positive direction and also minimizes the motor torque under the motor speed constraint:
solving $\vartheta(0,x)=\lim_{\epsilon\rightarrow0}\vartheta^\epsilon(0,x)$ where 
\begin{align}
    \vartheta^\epsilon (0,x)=\inf_{\alpha} \int_0^{1} \alpha(s,2) ds + 1000 \mathrm{x}(1,3) \label{eq:gearOpt_1}\\
    \quad \text{subject to } \begin{cases}
        \eqref{eq:ex_grear_Dyn1} \text{ holds}, \\
        \mathrm{x}(0,\cdot)=[0,0,0,0],\\
        \alpha(s,1) \in \{1,2 \},~~\alpha(s,2) \in [0, 1], \\
        \lvert \mathrm{x}(s,2) \rvert \leq 0.1 + \epsilon,
    \end{cases} 
    \label{eq:gearOpt_2}
\end{align}
for all $s\in [0,1]$.
The generalized Lax formula with the temporal discretization ($t_0=0,...,t_K=1$) provides a convex problem since Condition \ref{cvxCondition:ConvexCondition_HopfLax} is satisfied, however, the temporally discretion of the gear problem as in (\eqref{eq:Numerical_Problem_Cost} subject to \eqref{eq:Numerical_Problem_Const}) is non-convex.

To get the control constraint for the generalized Lax formula, we first derive $B(s,\mathrm{x}(s))$.
By the definition of $B(s,\mathrm{x}(s))$ in \eqref{eq:Trans_CtrlSet},
\begin{align*}
    B(s,\mathrm{x}(s)) &= \{\beta(s,1)=-\mathrm{x}(s,2), \beta(s,3)=-\mathrm{x}(s,4)\} \cap \\
    & \big( \{ \beta(s,2)=-\beta(s,4), \beta(s,2)\in[-\frac{1}{4},0]  \}  \\
    & \cup \{2\beta(s,2)=-\beta(s,4),\beta(s,2)\in[-\frac{1}{13 },0] \} \big).
\end{align*}
As described in Figure \ref{fig:Example_gear} (b), $B(s,\mathrm{x}(s))$ is not convex but $\text{Conv}(B(s,\mathrm{x}(s)))$ is convex.
\begin{align*}
    &\text{Conv}(B(s,\mathrm{x}(s)))= \{
    \beta(s,1)=-\mathrm{x}(s,2), \beta(s,3)=-\mathrm{x}(s,4),\\
    &\quad\quad\quad\quad\quad\beta(s,2)+\beta(s,4) \leq 0, -\beta(s,2)-2\beta(s,4) \leq 0,\\
    &\quad\quad\quad\quad\quad 5\beta(s,2) + 9\beta(s,4) \leq 1  \}
\end{align*}

We will derive the stage cost ($H^*$) for the generalized Lax formula in \eqref{eq:StateConst_Trans_ValueFunc} by Lemma \ref{lemma:Lemma1}.
If $\alpha(s,1)=1$, then $L^b (s,\mathrm{x}(s),\beta(s))= 5\beta(s,2)+9\beta(s,4)$ since $\beta(s,4)=-\beta(s,2)$.
If $\alpha(s,1)=2$, then $L^b (s,\mathrm{x}(s),\beta(s))=5\beta(s,2)+9\beta(s,4)$ since $\beta(s,4)=-2\beta(s,2)$.
Since $5\beta(s,2)+9\beta(s,4)$ is affine in $(\beta(s,2),\beta(s,4))$,
\begin{align}
    H^*(s,\mathrm{x}(s),\beta(s)) =5\beta(s,2)+9\beta(s,4)
    \label{eq:ex_stageCost_Same}
\end{align}
by Lemma \ref{lemma:Lemma1}.

The generalized Lax formula in Theorem \ref{thm:Value_Equal_StateConstraint} provides the following optimal control problem:
\begin{align}
    &\quad\quad\inf_{\beta} \int_0^{1} 5 \beta(s,2)+9 \beta (s,4) ds + 1000 \mathrm{x}(1,3) \label{eq:Gear_GHL_cost}\\
    &\quad \text{subject to } \begin{cases}
        \dot{\mathrm{x}} (s)=-\beta(s), \\
        \mathrm{x}(0,\cdot)=[0,0,0,0],\\
        \beta (s,1) = -\mathrm{x}(s,2), ~~
        \beta (s,3) = -\mathrm{x}(s,4), \\
        -\beta(s,2) - \beta(s,4) \leq 0,\\
        2\beta(s,2) + \beta(s,4) \leq 0, \\
        5 \beta(s,2) + 9 \beta(s,4) \leq 1,\\
        \lvert \mathrm{x}(s,2) \rvert \leq 0.1,
    \end{cases} 
    \label{eq:Gear_GHL_const}
\end{align}
for all $s\in [0,1]$. This is a convex problem.
We observe that the state-constrained optimal control problem (\eqref{eq:gearOpt_1} subject to \eqref{eq:gearOpt_2}) contains the discrete constraint, which is a non-convex constraint. 
On the other hand, the constraints of the generalized Lax formula are convex.

The generalized Lax formula (\eqref{eq:Gear_GHL_cost} subject to \eqref{eq:Gear_GHL_const}) is numerically solved by the interior-point method \cite{boyd2004convex}, and we denote the optimal state sequence $\mathrm{x}_*[\cdot]$ and control sequence $\beta_*[\cdot]$.

Algorithm \ref{alg:Opt_NewFormulation1} numerically computes an optimal control signal $\alpha_*^\epsilon$, and Figure \ref{fig:Example_gear} (b) explains how to linearly decompose the control $b\in\text{Conv}(B(s,\mathrm{x}(s)))$ into a finite $b_i$ in $B(s,\mathrm{x}(s))$.
By substituting $t_k$ for $s$ and $\mathrm{x}_*[k]$ for $\mathrm{x}(s)$ in Figure \ref{fig:Example_gear} (b), we have
\begin{align*}
    \beta_*[k] =  \gamma_{1}[k] \beta_{1*} [k] + \gamma_{2}[k] \beta_{2*}[k]
\end{align*}
for $\beta_*[k]\in\text{Conv}(B (t_k,\mathrm{x}_*[k]))$ for some $\gamma_{1*}[k],\gamma_{2*}[k]\in[0,1]$ such that $\gamma_{1*}[k]+\gamma_{2*}[k]=1$.
The mathematical expression for $\beta_{i*}[k]$ is in Figure \ref{fig:Example_gear} (b).
Note that \eqref{eq:LinearComb_cost} also holds for $\beta_{i*}[k]$ and $\gamma_{i*}(s)$ by \eqref{eq:ex_stageCost_Same}.
Then, we can compute
\begin{align}
    \alpha^{\epsilon }_*(s) = \begin{cases}\alpha_{1*}[k],&s\in[t_k,t_k+\Delta t_k \gamma_1[k]),\\\alpha_{2*}[k],&s\in[t_k+\Delta t_k \gamma_1[k],t_{k+1}),\end{cases}
\end{align}
where $\alpha_{1*}[k]$ and $\alpha_{2*}[k]$ satisfy
\begin{align}
    \begin{split}
        &\beta_{1*}[k]=-f(t_k,\mathrm{x}_{*}[k],\alpha_{1*}[k]),\\
        &\beta_{2*}[k]=-f(t_k,\mathrm{x}_{*}[k],\alpha_{2*}[k]),
    \end{split}
\end{align}
and $\mathrm{x}_*^{\epsilon}$ solving \eqref{eq:ex_grear_Dyn1} for $\alpha^{\epsilon}_*$. Note that $\mathrm{x}_*[\cdot]$ is the optimal state sequence for the temporally discretized generalized Lax formula \eqref{eq:Gear_GHL_cost} subject to \eqref{eq:Gear_GHL_const}.

The numerical results using the generalized Lax formula and Algorithm \ref{alg:Opt_NewFormulation1} are shown in Figure \ref{fig:Example_gear_result}.
Figure \ref{fig:Example_gear_result} (a) shows an optimal state trajectory ($\mathrm{x}_*^\epsilon$), and Figure (b) shows an optimal control signal ($\alpha_*^\epsilon$).
As we observed in the previous two examples, we have a frequent control switching in this example. 
This issue is discussed in the later part of Section \ref{sec:NewFormulation} where we provide practical suggestions to overcome this issue. 
In this paper, we do not provide theoretical support to validate these suggestions: this is one of our future work.
The computation time for the generalized Lax formula is 6.36 s.

For this gear system problem, non-convex programming such as MIP is not necessary. Instead, the generalized Lax formula provides a convex problem.


\section{Conclusion and Future Work} 
\label{sec:conclusion}

This paper proposes a computationally efficient Hamilton-Jacobi-based formula (\textit{the generalized Lax formula}) for the finite-horizon state-constrained optimal control problem, and presents a numerical algorithm to compute optimal control signal and state trajectory. 
It is proved that the optimal cost and state trajectory computed by the proposed method converge to the optimal solutions as the size of the temporal discretization converges to zero.
Our method is computationally efficient to provide an optimal solution if the dynamics are affine in the state, and the stage and terminal costs, as well as the state constraint, are convex in the state but not necessarily in the control, as described in Condition \ref{cvxCondition:ConvexCondition_HopfLax}.
Three practical examples show the computational efficiency of our method: the generalized Lax formula and the numerical algorithm. 
Our method requires 152 s to obtain a solution of the 12D problem, for which it is not realistic to utilize the grid-based methods, such as the level-set method and the fast marching method, due to the exponential complexity in computation.


Our future directions are 1) to find a method for an optimal control signal that has less control switching than our method, 2) to extend our generalized Lax formula for the time-optimal problems \cite{mitchell2005time} and infinite-horizon optimal control problems \cite{Bertsekas2015Book}, 3) to extend the theory in this paper for the state-constrained optimal control problem with discrete states, and 4) to derive a Hamilton-Jacobi-based formula for two-player games \cite{evans1984differential}.

\input{Appendix.tex}
\section*{Acknowledgements}
The authors thank Ellis Ratner, Sang Min Han, and Margaret P. Chapman for discussions.


\addtolength{\textheight}{0cm}   

\bibliographystyle{IEEEtran}
\bibliography{IEEEexample}

\vspace*{-2\baselineskip}
\begin{IEEEbiography}
    [{\includegraphics[width=1in,height=1.25in,clip,keepaspectratio]{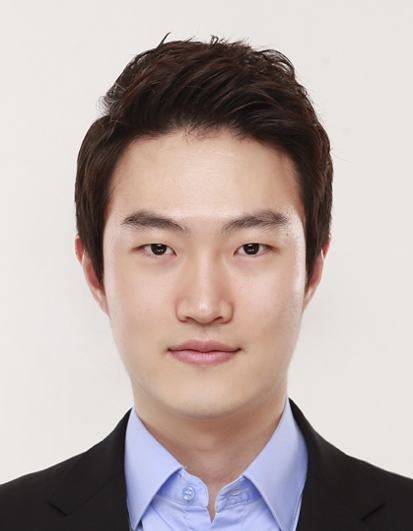}}]{Donggun Lee}
is a Ph.D. student in Mechanical Engineering at UC Berkeley. He received B.S. and M.S. degrees in Mechanical Engineering from Korea Advanced Institute of Science and Technology (KAIST), Daejeon, Korea, in 2009 and 2011, respectively. Donggun works in the area of control theory and robotics.
\end{IEEEbiography}
\vspace*{-2\baselineskip}
\begin{IEEEbiography}
    [{\includegraphics[width=1in,height=1.25in,clip,keepaspectratio]{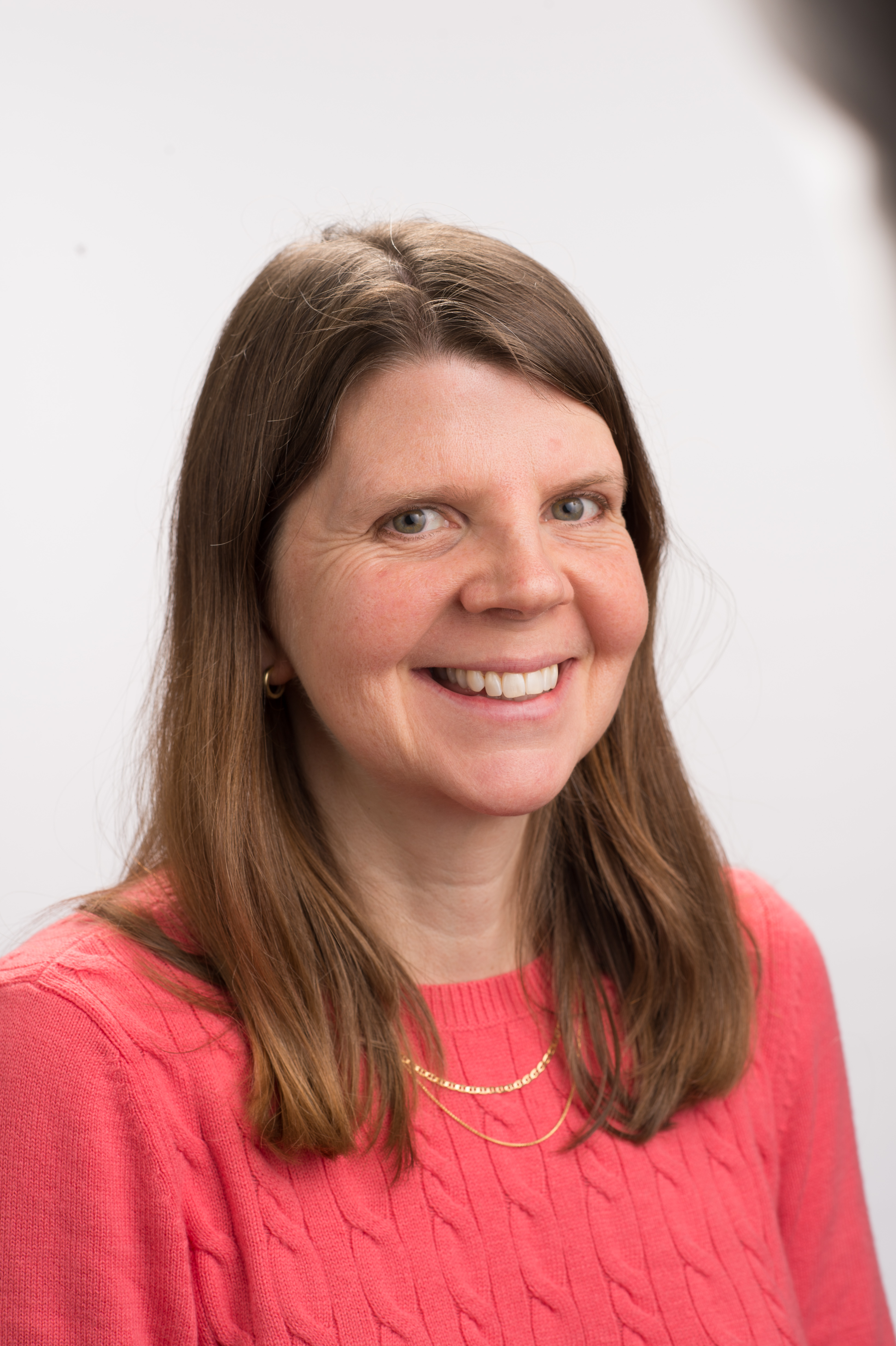}}]{Dr. Claire Tomlin}
is the Charles A. Desoer Professor of Engineering in EECS at Berkeley. She was an Assistant, Associate, and Full Professor in Aeronautics and Astronautics at Stanford from 1998 to 2007, and in 2005 joined Berkeley. Claire works in the area of control theory and hybrid systems, with applications to air traffic management, UAV systems, energy, robotics, and systems biology. She is a MacArthur Foundation Fellow (2006), an IEEE Fellow (2010), in 2017 she was awarded the IEEE Transportation Technologies Award, and in 2019 was elected to the National Academy of Engineering and the American Academy of Arts and Sciences.
\end{IEEEbiography}

\end{document}

%% file: Appendix.tex
\appendix

\subsection{Proof of Lemma \ref{lemma:Lemma1}}
\label{appen:Lemma1_proof}
\noindent
(i) Proof of \eqref{eq:property_ConvConj0} and \eqref{eq:property_ConvConj}

The corresponding Hamiltonian for $\vartheta$ in \eqref{eq:ValueFunction_Def_CtrlTrans} has to be the same with the Hamiltonian in \eqref{eq:Hamiltonian}:
\begin{align*}
    \max_{b\in B(s,x)}[ p \cdot b -L^b (s,x,b) ] = H(s,x,p),
\end{align*}
since $\vartheta$ in \eqref{eq:ValueFunction_Def_CtrlTrans} subject to \eqref{eq:ValueFunction_Constraint_CtrlTrans} is the unique viscosity solution to the HJB PDE in Theorem \ref{thm:HJB_PDE}.
Hence, $H\equiv (L^b)^*$ and $H^*\equiv (L^b)^{**}$
since $L^b$ is semi lower-continuous in $b$ for each $(s,x)\in[t,T]\times\R^n$.

\noindent
(ii) Proof of \eqref{eq:Dom_ConvCtrl}

\noindent
Case 1. $b\in \text{Conv}(B(s,x))$ 

There exist $b_i \in B(s,x)$ and $\gamma_i \geq 0$ ($\sum_i \gamma_i =1$) such that $b =\sum_i \gamma_i b_i$.
Since $H^*(s,x,\cdot)$ is convex in $b$,
\begin{align*}
    H^* (s,x,b) \leq \sum_i \gamma_i H^* ( s,x,b_i) < \infty
\end{align*}
since each of $H^*( s,x,b_i)$ is finite.
\\\\
\noindent
Case 2. $b\notin \text{Conv}(B(s,x))$

For two closed convex sets $\{b\}$ and $\text{Conv}(B(s,x))$, the seperating hyperplane theorem \cite{boyd2004convex} implies that there exists a hyperplane ($P:\R^n \rightarrow \R$): $P (b') := p' \cdot b' + c$ such that
\begin{align*}
    P (b') \begin{cases}>0,&b'\notin \text{Conv}(B(s,x)) \\ \leq 0, & b' \in \text{Conv}(B(s,x)).\end{cases}
\end{align*}
By picking $p= \gamma p'$ where $\gamma\in\R$,
\begin{align*}
    H^* ( s, x,b) &=\max_p \min_{b'} [p\cdot (b-b') +L^b (s,x,b')]\\
    & \geq \max_{\gamma} \min_{b'} \gamma p' \cdot (b-b') +L^b (s,x,b')=\infty 
\end{align*}
since $p'\cdot(b-b')>0$ for all $b'\in\text{Conv}(B(s,x))$.
$\qed$


\subsection{Additional Lemma used in the proof of Theorem \ref{thm:OptCtrl_PostProcess}}
\label{appen:Lemma_TrajBounded_proof}
In this appendix, we describe the below lemma that is used in the proof of Theorem \ref{thm:OptCtrl_PostProcess}.
\begin{lemma}
    Suppose Assumption \ref{assum:BigAssum} holds.
    Given the initial state $x\in\R^n$, there exists a constant $C>0$ such that
    \begin{align}
        \lVert \mathrm{x}(s)-x \rVert \leq C (s-t)
    \end{align}
    for all $s\in[t,T]$, $\alpha\in \mathcal{A}(t)$ where $\mathrm{x}$ solves \eqref{eq:NoConst_vartheta_const}.
    \label{lemma:lemma_TrajBounded}
\end{lemma}

\noindent
\textbf{Proof.} For any control $\alpha\in\mathcal{A}(t)$ and $\mathrm{x}$ solving \eqref{eq:NoConst_vartheta_const}, 
\begin{align*}
    &\| \mathrm{x}(s) - x \| =\| \int_t^s f(\tau ,\mathrm{x}(\tau),\alpha(\tau))d\tau\| \\
    \leq & \int_t^s \lVert f(\tau,\mathrm{x}(\tau),\alpha(\tau)) -f(t,x,\alpha(\tau))\rVert + \| f(t,x,\alpha(\tau)) \rVert d\tau\\
    \leq & c' + L_f \int_t^s \| \mathrm{x}(\tau)-x\| d\tau \text{ by Lipchitz continuity of } f,
\end{align*}
where $c'=\frac{1}{2}L_f (s-t)^2 + \max_{a\in A }\{f(t,x,a)\}(s-t)$.
By the Gronwall's inequality, 
\begin{align*}
    \| \mathrm{x}(s) -x \| &\leq c'+c' \exp(L_f (s-t))\leq C(s-t),
\end{align*}
where $C = [\frac{1}{2}L_f (T+t)+\max_{a\in A}\{f(t,x,a)\}](1+\max\{1,\exp(L_f (T-t))\})$.
$\qed$

\subsection{Proof of Theorem \ref{thm:OptCtrl_PostProcess}}
\label{appen:Theorem_OptCtrl}

\noindent
\textbf{Proof.} \\
\textbf{Step 1.}
For a temporal partition $t_0=t<...<t_K = T$, consider a state trajectory ($\mathrm{x}_0 :[t,T]\rightarrow \mathbb{R}^n$) solving
\begin{align*}
    \dot{\mathrm{x}}_0^\epsilon (s) = -\beta(t_k), \quad s\in[t_k,t_{k+1})
\end{align*}
for $s\in[t,T]$ and $\mathrm{x}_0(t)=x$.
Since $\beta$ is Riemann integrable, for all $\epsilon$, there exists $\delta >0$ such that $|t_{k+1} - t_k| <\delta$ for all $k$ and 
\begin{align}
    \| \mathrm{x}_0 - \mathrm{x} \|_{L^\infty (t,T)} < \epsilon.
    \label{eq:proof_thm5_1}
\end{align}
As $\delta$ converges to 0, $\epsilon$ converges to 0.

Since $\beta (t_k)$ is in $\text{Conv}(B(t_k ,\mathrm{x}(t_k)))$, by Lemma \ref{lemma:lemma6},
there exists $b_i^k  \in B(t_k,\mathrm{x}(t_k))$, $a_i^k \in A$, and $\gamma_i^k$ such that 
\begin{align}
    H^*(t_k,\mathrm{x}(t_k),\beta (t_k)) = \sum_i \gamma_i^k L^b ( t_k ,\mathrm{x}(t_k),b_i^k) \notag\\ 
     = \sum_i \gamma_i^k L (t_k ,\mathrm{x}(t_k),a_i^k ),
    \label{eq:Thm5_Proof_1_1}\\
    \beta(t_k) = \sum_i \gamma_i^k b_i^k,\quad
    b_i^k = -f(t_k, \mathrm{x}(t_k), a_i^k),
    \label{eq:Thm5_Proof_1_2}
\end{align}
and $\sum_i \gamma_i^k = 1$ for $k=0,...,K-1$.

\noindent
\textbf{Step 2.} In each time interval $[t_k,t_{k+1})$, we construct a finer temporal discretization: $[t_{k,0}=t_k,...,t_{k,i},...,t_{k,\bar{i}_k}=t_{k+1})$, where $t_{k,i} = t_k + \sum_{j=1}^{i}\gamma_j^k \Delta t_k$ and $\bar{i}_k$ denotes the number of $\gamma_i^k$ in \eqref{eq:Thm5_Proof_1_1} for each $k$.
Define a control input $\beta_1:[t,T]\rightarrow\R^n$:
\begin{equation*}
    \beta_1(s) = b_{i+1}^k ,\quad s\in [t_{k,i}, t_{k,i+1}),
\end{equation*}
where $b_i^k$ is defined in \eqref{eq:Thm5_Proof_1_1} and \eqref{eq:Thm5_Proof_1_2}, and also define the corresponding state trajectory ($\mathrm{x}_1:[t,T]\rightarrow \mathbb{R}^n$) solving
\begin{align*}
    \dot{\mathrm{x}}_1 (s) &= -\beta_1 (s) = f(t_k, \mathrm{x}(t_k), a_{i+1}^k )
\end{align*}
for $s \in [t_{k,i},t_{k,i+1})$, and $\mathrm{x}_1(t)=x$.
Then, for $k=0,...,K$, $\mathrm{x}_1 (t_{k})=\mathrm{x}_0 (t_{k})$, and for $s\in[t_k,t_{k+1})$,
\begin{align}
     \| \mathrm{x}_1 (s) -  \mathrm{x}(s) \|  &\leq \| \mathrm{x}_1 (s) - \mathrm{x}_0 (s) \| +\| \mathrm{x}_0 (s) - \mathrm{x}(s) \|
    \notag\\
    &\leq   \int_{t_k}^{s} \| -\beta_1(\tau) + \beta(t_k)\rVert  d\tau  + \epsilon \quad \text{by }\eqref{eq:proof_thm5_1}\notag\\
    &\leq  c_1 \delta  + \epsilon,
    \label{eq:Thm2Proof_Step2}
\end{align}
where $c_1 =\max_{k,i}\lVert b_i^k -\beta(t_k) \rVert$. $c_1$ is bounded since $\mathrm{x}$ and $A$ are bounded.


\noindent
\textbf{Step 3.} 
We consider a control input $\alpha_2\in\mathcal{A}(t)$
\begin{align}
    \alpha_2(s) = a_{i+1}^k,\quad s\in [t_{k,i}, t_{k,i+1}),
    \label{eq:proof_ctrl1}
\end{align}
where $a_i^k$ is defined in \eqref{eq:Thm5_Proof_1_1} and \eqref{eq:Thm5_Proof_1_2}.
and the corresponding state trajectory ($\mathrm{x}_2:[t,T]\rightarrow\mathbb{R}^n$) solving
\begin{align*}
    \dot{\mathrm{x}}_2(s) &= f(s, \mathrm{x}_2(s), \alpha_2(s) ), \quad s\in[t,T].
\end{align*}
We have
\begin{align}
    & \lVert \mathrm{x}_2 (t_{k,i+1})-\mathrm{x}_1(t_{k,i+1}) \rVert \notag\\
    \leq & \lVert \mathrm{x}_2 (t_{k,i})-\mathrm{x}_1 (t_{k,i}) \rVert + \int_{t_{k,i}}^{t_{k,i+1}} \lVert f(\tau,\mathrm{x}_2(\tau),a_{i+1}^k ) \notag\\
     & \quad\quad\quad\quad\quad\quad\quad\quad\quad\quad\quad\quad -f(t_k, \mathrm{x}(t_k ), a_{i+1}^k )  \rVert d\tau \notag\\
    \leq & \lVert \mathrm{x}_2 (t_{k,i})-\mathrm{x}_1(t_{k,i}) \rVert + \int_{t_{k,i}}^{t_{k,i+1}} L_f \big( (\tau-t_{k})\notag\\
    + & \lVert \mathrm{x}_2 (\tau)-\mathrm{x}_2 (t_{k,i}) \rVert + \lVert \mathrm{x}_2 (t_{k,i})  - \mathrm{x}_1 (t_{k,i} ) \rVert \notag\\
    +& \lVert \mathrm{x}_1 (t_{k,i})-\mathrm{x}_1 (t_k) \rVert + \lVert \mathrm{x}_1 (t_k) -\mathrm{x}(t_k)\rVert \big) d\tau \label{eq:Thm5Proof_Step3}
\end{align}
by Lipschitz of $f$.

By the way, for $\tau\in[t_{k,i},t_{k,i+1})$,
\begin{align}
    \lVert \mathrm{x}_2 (\tau)-\mathrm{x}_2 (t_{k,i})  \rVert \leq c_2 \delta
    \label{eq:Thm5Proof_Step3_2}
\end{align}
for some $c_2$ by Lemma \ref{lemma:lemma_TrajBounded}, and
\begin{align}
    \lVert \mathrm{x}_1 (t_{k,i}) -\mathrm{x}_1 (t_k)\rVert& =\bigg\| \sum_{j=1}^{i} f(t_k, \mathrm{x} (t_k),a_{j}^k) (t_{k,j}-t_{k,j-1})  \bigg\|\notag\\
    &\leq c_3\delta
    \label{eq:Thm5Proof_Step3_3}
\end{align}
for some $c_3$ since $f(t_k,\mathrm{x}(t_k),a_j^k)$ is bounded.

By \eqref{eq:Thm5Proof_Step3_2}, \eqref{eq:Thm5Proof_Step3_3}, and \eqref{eq:Thm2Proof_Step2}, \eqref{eq:Thm5Proof_Step3} becomes
\begin{align*}
    &\| \mathrm{x}_2 (t_{k,i+1})-\mathrm{x}_1(t_{k,i+1}) \| \\
    \leq& (1+L_f (t_{k,i+1}-t_{k,i}))\| \mathrm{x}_2 (t_{k,i})-\mathrm{x}_1(t_{k,i}) \| \\
    +& L_f (t_{k,i+1}-t_{k,i}))(\delta c_4  + \epsilon)
\end{align*}
for some $c_4 >0$.
This is equivalent to
\begin{align}
    & \lVert \mathrm{x}_2 (t_{k,i+1})-\mathrm{x}_1 (t_{k,i+1}) \rVert + \delta c_4+\epsilon\notag\\
    \leq & \big(1+L_f (t_{k,i+1}-t_{k,i})\big)(\lVert \mathrm{x}_2 (t_{k,i})-\mathrm{x}_1 (t_{k,i}) \rVert + \delta c_4+\epsilon).
    \label{eq:Thm5Proof_Step3_4}
\end{align}
By multiplying the both side of \eqref{eq:Thm5Proof_Step3_4} for all $k'\geq 0,i'\geq 0$ such that $t \leq t_{k',i'}\leq t_{k,i}$, we have
\begin{align}
    &\lVert \mathrm{x}_2 (t_{k,i})-\mathrm{x}_1 (t_{k,i}) \rVert \notag\\
    \leq & \bigg[\prod_{k',i'}^{t_{k',i'}\leq t_{k,i}} (1+L_f (t_{k',i'}-t_{k',i'-1})) -1\bigg](\delta c_4+\epsilon)\notag\\
    \leq & \bigg[\prod_{k',i'}^{t_{k',i'}\leq T} (1+L_f (t_{k',i'}-t_{k',i'-1})) -1\bigg](\delta c_4+\epsilon)\notag\\
    \leq & \bigg[ \big(1+ \frac{L_f T}{\bar{m}}\big)^{\bar{m}} -1\bigg](\delta c_4+\epsilon)
    \label{eq:Thm5Proof_Step3_5}
\end{align}
for $\bar{m}=\sum_{k',i'}1$ subject to $t\leq t_{k',i'}\leq T$: the number of the discrete time points in $[t,T]$. 
The last inequality in \eqref{eq:Thm5Proof_Step3_5} holds since 
$\prod_{i=1}^n x_i \leq (\frac{\sum_{i=1}^n x_i }{n})^n$ for any positive $x_i$s and $n\in\mathbb{N}$.
Note that $\bar{m}\rightarrow\infty$ as $\delta\rightarrow0$, and 
\begin{align*}
    \lim_{\bar{m}\rightarrow \infty} \Big(1+\frac{L_f T}{\bar{m}}\Big)^{\bar{m}} = \exp(L_f T)<\infty.
\end{align*}
Thus, \eqref{eq:Thm5Proof_Step3_5} implies
\begin{align}
    \lVert \mathrm{x}_2 (t_{k,i}) - \mathrm{x}_1 (t_{k,i}) \rVert \leq c_5 \delta + c_6 \epsilon
    \label{eq:Thm5Proof_Step3_6}
\end{align}
for all $t_{k,i}\in[t,T]$.

For $s\in(t_{k,i},t_{k,i+1})$,
\begin{align*}
    &\lVert \mathrm{x}_2 (s)-\mathrm{x}_1 (s) \rVert \\
    \leq & \lVert \mathrm{x}_2(s)-\mathrm{x}_2(t_{k,i}) \rVert + \lvert \mathrm{x}_2(t_{k,i})-\mathrm{x}_1(t_{k,i})\rVert \\
    +&\lVert \mathrm{x}_1 (t_{k,i})-\mathrm{x}_1(s)\rVert\\
    \leq & c_7 \delta +c_8\epsilon
\end{align*}
for some $c_7,c_8>0$ by \eqref{eq:Thm5Proof_Step3_2}, \eqref{eq:Thm5Proof_Step3_6}, and \eqref{eq:Thm5Proof_Step3_3} with little modification.

\noindent
\textbf{Step 4.} We choose $\mathrm{x}^\epsilon=\mathrm{x}_2$ and $\alpha^\epsilon=\alpha_2$ for approximate state and control trajectories in Theorem \ref{thm:OptCtrl_PostProcess}. To sum up Step 1 to Step 3, 
\begin{align}
    \lVert \mathrm{x}^\epsilon(s) -\mathrm{x}(s)\rVert \leq c_9 \delta + c_{10}\epsilon, \quad s\in[t,T].
    \label{eq:Thm5Proof_4}
\end{align}
As $\delta \rightarrow 0$, $\epsilon \rightarrow 0$ and $\|\mathrm{x}^\epsilon-\mathrm{x}\|_{L^\infty (t,T)} \rightarrow 0$.
This proves \eqref{eq:OptTraj_approximation}.

\noindent
\textbf{Step 5.} 
Define a discrete sum of $H^*$:
\begin{align*}
    S (t) &\coloneqq \sum_{k=0}^{K-1} H^* (t_{k}, \mathrm{x}(t_{k}),\beta(t_{k}))(t_{k+1}-t_{k}) .
\end{align*}
Note that $K$ denotes the index of the discrete time point for the terminal time: $t_K = T$.
Since $H^* (s,\mathrm{x}(s),\beta(s))$ is also Riemann integrable, 
\begin{align}
    \bigg\lvert \int_t^T H^* (s,\mathrm{x}(s),\beta(s)) ds - S (t) \bigg| < \epsilon_1
    \label{eq:Thm5_Proof_5_0}
\end{align}
for some $\epsilon_1>0$, and $\lim_{\delta \rightarrow 0}\epsilon_1=0$ where $\delta \geq\max_i \Delta t_k$ ($k=0,...,K-1$).

By \eqref{eq:Thm5_Proof_1_1}, \eqref{eq:Thm5_Proof_1_2}, \eqref{eq:Thm5Proof_4}, and Lipschitz continuity of $L$ in Assumption \ref{assum:BigAssum},
\begin{align}
    &\bigg\lvert \int_{t_k}^{t_{k+1}} L(s,\mathrm{x}^\epsilon(s),\alpha^\epsilon (s)) -H^*(t_k,\mathrm{x}(t_k),\beta(t_k)) ds\bigg\rvert \notag\\
     =& \bigg\lvert\sum_i \int_{t_{k,i}}^{t_{k,i+1}}L(s,\mathrm{x}^\epsilon(s),a_{i+1}^k)-L(t_k, \mathrm{x}(t_k),a_{i+1}^k)ds \bigg\lvert \notag\\
     \leq & \Delta t_k L_L (c_{11} \delta + c_{12}\epsilon)
    \label{eq:Thm5_Proof_5_1}
\end{align}
for some $c_{11},c_{12}>0$, where $L_L$ is the Lipschitz constant of $L$.
Therefore,
\begin{align}
    &\bigg\lvert \int_{t}^{T} L(s,\mathrm{x}^\epsilon(s),\alpha^\epsilon (s))ds -S(t)\bigg\rvert \leq TL_L (c_{11} \delta + c_{12}\epsilon).
    \label{eq:Thm5_Proof_5_2}
\end{align}

By \eqref{eq:Thm5_Proof_5_0}, \eqref{eq:Thm5_Proof_5_2}, \eqref{eq:Thm5Proof_4}, and Lipschitz continuity of $g$ in Assumption \ref{assum:BigAssum},
\begin{align*}
    &\bigg| \int_t^T H^*(s,\mathrm{x}(s),\beta(s))ds +g(\mathrm{x}(T))\\
    &\quad\quad\quad\quad -\int_t^T L(s,\mathrm{x}^\epsilon(s),\alpha^\epsilon(s))ds-g(\mathrm{x}^\epsilon(T))\bigg|\\
    &\quad\quad <c_{13}\delta + c_{14}\epsilon + \epsilon_1,
\end{align*}
where $c_{13} = TL_L c_{11} + L_g c_9$, $c_{14}=TL_L c_{12} + L_g c_{10}$, where $L_g$ is the Lipschitz constant of $g$.
This proves \eqref{eq:OptCtrl_approximation}. $\qed$